\documentclass[aps,prx,groupedaddress,nofootinbib,notitlepage,showpacs,floatfix,twocolumn]{revtex4-1}

\usepackage{amsmath,amssymb,setspace}
\usepackage{IEEEtrantools}
\usepackage{physics}
\usepackage{calc}
\usepackage{dsfont}
\newcommand\numberthis{\addtocounter{equation}{1}\tag{\theequation}}

\usepackage{graphicx}
\usepackage[usenames]{color}
\usepackage{bm, bbm} 
\usepackage{color}
\usepackage{srcltx}
\usepackage{hyperref}
\usepackage{slashed}
\hypersetup{
    colorlinks=true, 
    linkcolor=cyan, 
    citecolor=magenta, 
    filecolor=magenta, 
    urlcolor=cyan, 
    runcolor=cyan
}
\usepackage{threeparttable}
\usepackage{upgreek}
\usepackage{marginnote}
\usepackage{subfigure}
\usepackage{soul}

\newcommand{\bes}{\begin{subequations}}
\newcommand{\ees}{\end{subequations}}
\newcommand{\bea}{\begin{eqnarray}}
\newcommand{\eea} {\end{eqnarray}}
\newcommand{\beq}{\begin{equation}}
\newcommand{\eeq}{\end{equation}}

\newcommand{\splu}{S_{+}}

\renewcommand{\to}{\rightarrow}

\newcommand{\ud}{\dd}
\newcommand{\odf}{\dv}

\def\Tr{\mathrm{Tr}}
\DeclareMathOperator{\erfc}{erfc}

\begin{document}
\title{A double-slit proposal for quantum annealing}
\author{Humberto Munoz-Bauza$^{(1,4)}$}
\author{Huo Chen$^{(2,4)}$}
\author{Daniel Lidar$^{(1,2,3,4)}$}
	\affiliation{$^{(1)}$Department of Physics and Astronomy, University of Southern California, Los Angeles, California 90089, USA}
	\affiliation{$^{(2)}$Department of Electrical and Computer Engineering, University of Southern California, Los Angeles, California 90089, USA}
	\affiliation{$^{(3)}$Department of Chemistry, University of Southern California, Los Angeles, California 90089, USA}
	\affiliation{$^{(4)}$Center for Quantum Information Science \&
		Technology, University of Southern California, Los Angeles, California 90089, USA}

\begin{abstract}
We formulate and analyze a double-slit proposal for quantum annealing, which involves observing the probability of finding a two-level system (TLS) undergoing evolution from a transverse to a longitudinal field in the ground state at the final time $t_f$. We demonstrate that for annealing schedules involving two consecutive diabatic transitions, an interference effect is generated akin to a double-slit experiment. The  observation of oscillations in the ground state probability as a function of $t_f$ (before the adiabatic limit sets in) then constitutes a sensitive test of coherence between energy eigenstates. This is further illustrated by analyzing the effect of coupling the TLS to a thermal bath: increasing either the bath temperature or the coupling strength results in a damping of these oscillations. The theoretical tools we introduce significantly simplify the analysis of the generalized Landau-Zener problem. Furthermore, our analysis connects quantum annealing algorithms exhibiting speedups via the mechanism of coherent diabatic transitions to near-term experiments with quantum annealing hardware. 
\end{abstract}

\maketitle

\section{Introduction}
Feynman famously wrote that the double-slit interference experiment ``...has in it the heart of quantum mechanics. In reality, it contains the only mystery''~\cite{Feynman-vol3}. Here we propose a double-slit experiment for quantum annealing (QA). In analogy to Feynman's particle-wave double-slit, the proposed experiment can only be explained by the presence of interference and would break down upon either an intermediate measurement or strong decoherence.
We are motivated by the recent resurgence of interest in quantum annealing using the transverse field Ising model~\cite{kadowaki_quantum_1998,RevModPhys.80.1061}, which has led to major efforts to build physical quantum annealers for the purpose of solving optimization and sampling problems~\cite{Dwave,Weber:2017aa,Quintana:2017aa,Novikov:2018aa}, and significant debate as to whether quantum effects are at play in the performance of such devices~\cite{q108,SSSV}. The mechanisms by which QA might achieve a speedup over classical computing remain hotly contested, and while tunneling is often promoted as a key ingredient~\cite{PhysRevX.6.031015} and entanglement is often viewed as a necessary condition which must be demonstrated~\cite{DWave-entanglement,Albash:2015pd}, a consensus has yet to emerge. 
Yet, an explicit example is known where QA theoretically provides an oracle-based exponential quantum speedup over all classical algorithms~\cite{Somma:2012kx}, and other examples are known where QA provides a speedup over classical simulated annealing~\cite{Farhi-spike-problem,2014arXiv1410.8484C,Muthukrishnan:2015ff,2015arXiv151106991K,Brady:2016aa,Jiang:2017aa}. An essential feature in all these cases are diabatic transitions which circumvent adiabatic ground state evolution to enable the speedup, in the spirit of the idea of shortcuts to adiabaticity~\cite{Campo:2013ix}. When these transitions result in a coherent recombination of the ground state amplitude (a phenomenon known as a diabatic cascade~\cite{Muthukrishnan:2015ff,Brady:2017aa}), 
the result is a wave-like interference pattern in the ground state probability as the anneal time is varied~\cite{Wiebe:12,Wecker:2016lh,Brady:2018aa}.
We thus conjecture that coherent recombination of ground state amplitudes after coherent evolution between diabatic transitions can play a critical role in enabling quantum speedups in QA. The double-slit proposal we formulate and analyze here is designed to test for the presence of quantum interference due to such coherent evolution. 

Viewed from a different perspective, our double-slit proposal joins a family of protocols designed to probe 
the dynamics of what Berry called the ``simplest non-simple quantum problem"~\cite{BERRY:1995aa}, a driven TLS near level crossings~\cite{Grifoni:1998aa}.
The two-level paradigm was introduced long ago by Landau and Zener (LZ)~\cite{LZ1-1,LZ1-2}. The corresponding Hamiltonian for the generalized LZ problem is
\begin{equation}
H_S(t) = -a(t)X-b(t)Z\ ,
\label{eq:H_S(t)}
\end{equation}
where $X$, $Y$ and $Z$ are the Pauli matrices.
In the original protocol which LZ solved analytically, $a(t)$ is constant, $b(t)$ is linear in $t$, and $t$ runs from $-\infty$ to $\infty$. The problem has since been studied under numerous variations, including Landau-Zener-Stueckelberg interferometry where $b(t)$ is periodic~\cite{stueckelberg_ecg_1932,ashhab_landauzenerstueckelberg_2017}, the subject of various experiments~\cite{Oliver:2005aa,Sillanpaa:2006aa,Petta:2010aa}. 
Complete analytical solutions were limited until recently to certain particular functional forms of $b(t)$ with constant $a(t)$~\cite{Bambini:1981aa}, a finite-range linear schedule for both $a(t)$ and $b(t)$~\cite{Vitanov:1996uq}, and periodic $a(t)$ and $b(t)$~\cite{Bezvershenko:2011aa}. An analytical solution for general $b(t)$ but constant $a(t)$ was found in Ref.~\cite{Barnes:2012fk}, which was then extended to general (but implicitly specified) $a(t)$ as well~\cite{Barnes:2013aa,Messina:2014aa}.
Here we consider the case of general schedules $a(t)$ and $b(t)$, and develop a simple to interpret,  yet surprisingly accurate, low-order time-dependent perturbation theory approach, that allows us to identify a class of schedules exhibiting ``giant" (relative to linear schedules) interference oscillations of the ground state population as a function of the total annealing time. Our proposal should in principle be straightforward to implement using, e.g., flux qubits, and toward this end we also study the effects of coupling to a thermal environment.

The structure of this paper is as follows. In Sec.~\ref{sec:closed} we analyze the TLS quantum annealing problem in the closed system limit. We first transform to an adiabatic interaction picture and perform a Magnus expansion, which allows us to give a simple expression for the ground state probability in terms of the Fourier transform of a key quantity we call the angular progression. We then analyze both the LZ problem (with a linear schedule) and a new ``Gaussian angular progression" schedule which gives rise to large interference oscillations. We explain how these oscillations can be interpreted in terms of a double-slit experiment generating interference between ground state amplitudes. In Sec.~\ref{sec:open} we analyze the problem in the presence of coupling to a thermal environment. We consider the weak-coupling limit both without and with the rotating wave approximation, and find the range of coupling strengths and temperatures over which the interference oscillations are visible, using parameters relevant for superconducting flux qubits. We find a simple semi-empirical formula that accurately captures all our open-system simulation results in terms of three physically intuitive quantities: the oscillation period, rate of convergence to the adiabatic limit, and damping due to coupling to the thermal environment. We express all three are in terms of the input parameters of the theory. Conclusions and the implications of our results are discussed in Sec.~\ref{sec:conc}. A variety of supporting technical calculations and bounds are provided in the Appendix.

\section{Closed system analysis and results}
\label{sec:closed}

\subsection{Adiabatic interaction picture for two-level system quantum annealing}
We first consider the closed system setting. Consider a two-level system (TLS) quantum annealing Hamiltonian in the standard form~\eqref{eq:H_S(t)}, where the annealing schedules $a(t),b(t)\geq 0$ respectively decrease/increase to/from $0$ 
with time $t\in [0,t_f]$, where $t_f$ is the duration of the anneal. The schedules need not be monotonic, and our analysis thus includes ``reverse annealing"~\cite{Perdomo-Ortiz:2011fh,Chancellor:2016ys,King:2018aa,Ohkuwa:2018aa,Venturelli:2018aa} as a special case. The TLS can be a single qubit or the two lowest energy levels of a multi-qubit system separated by a large gap from the rest of the spectrum.
Key to our analysis is a series of transformations designed to arrive at a conveniently reparametrized interaction picture. First, we rewrite Eq.~\eqref{eq:H_S(t)} in the form
\begin{equation}
H_S(s) = -\frac{1}{2}E_0 [A(s) Z +  B(s)Y],
\label{eq:H_S-new}
\end{equation}
where $A(s)=2a(t)/E_0$ and $B(s)=2b(t)/E_0$ are dimensionless schedules parametrized by the dimensionless time $s=t/t_f$, and $E_0>0$ is the energy scale of the Hamiltonian. We have cyclically permuted the Pauli matrices for later convenience. The ground states of $H_S(0)$ and $H_S(1)$ are $\ket{0}$ and $\ket{-i}$, respectively.
Second, we parametrize the annealing schedules in the angular form
\beq
\label{eq:2}
A(s) = \Omega(s)\cos\theta(s),\quad
B(s) = \Omega(s)\sin\theta(s)\ ,
\eeq
where $\theta(0)=0$ and $\theta(1)=\pi/2$.
Under this parametrization the eigenvalues of $H_S(s)$ are $\pm E_0 \Omega(s)/2$, so the gap is $\Delta(s) = E_0 \Omega(s)$.
Thus, any non-trivial time-dependence of the gap is encoded in the time-dependence of $\Omega(s)$, which we refer to as the dimensionless gap.
The quantity 
\beq
\tau(s) \equiv  \int_0^s \ud s'  \Omega(s')
\label{eq:3}
\eeq
is the cumulative dimensionless gap.
Third, changing variables from $s$ to $\tau$ to absorb $\Omega(s)$, the system satisfies the Schr\"odinger equation
\beq
i\odf{}{\tau}\ket{\psi} =- \frac{1}{2}E_0 t_f [\cos\theta(\tau)Z+\sin\theta(\tau)Y]\ket{\psi}
\label{eq:SE}
\eeq
 (we work in $\hbar =1$ units throughout).
The Hamiltonian is diagonalized at each instant by the rotation $R_X(\theta)=e^{-i\theta X /2}$.
Thus, fourth, we change into the adiabatic frame~\cite{klarsfeld_magnus_1992,Nalbach:2014aa} with
$\ket{\psi_{\mathrm{ad}}}=R_X(\theta)\ket{\psi}$, yielding:
\beq
i\odf{}{\tau}\ket{\psi_{\mathrm{ad}}} =H_{\mathrm{ad}}\ket{\psi_{\mathrm{ad}}}, \quad H_{\mathrm{ad}}(\tau) \equiv \frac{1}{2}\left(\dv{\theta}{\tau}X - E_0 t_f Z\right).
\label{eq:psi-ad}
\eeq
We call $\dv{\theta}{\tau}$ the \emph{angular progression} of the anneal.  

Finally, we transform into the interaction picture with respect to the free Hamiltonian $H_0 = -E_0 t_f Z/2$ and its propagator $U_0(\tau) = e^{-iH_0 \tau}$. Letting $S_\pm = (X\pm iY)/2$ denote the spin raising and lowering operators we have $X_{\mathrm{I}}(\tau) = U_0^\dagger(\tau)X U_0(\tau) =e^{-i E_0 t_f \tau}\splu + \textrm{h.c.}$, and obtain
\begin{equation}
i\odf{}{\tau}\ket{\psi_{\mathrm{I}}} = H_{\mathrm{I}}(\tau) \ket{\psi_{\mathrm{I}}}, \quad H_{\mathrm{I}}(\tau)\equiv \lambda(\tau)X_{\mathrm{I}}(\tau)\ ,
\label{eq:intp}
\end{equation}
where $\ket{\psi_{\mathrm{I}}} = U_0^{\dag}\ket{\psi_{\mathrm{ad}}}$ and $\lambda(\tau) = \frac{1}{2}\dv{\theta}{\tau}$.
Therefore, we see that in this \emph{adiabatic interaction picture} the dynamics of the annealed TLS is a rotation about the time-dependent $X_{\mathrm{I}}$ axis with a rate equal to the angular progression.

\subsection{Magnus expansion}
The corresponding time-ordered propagator $U_{\mathrm{I}}(\tau) = T_{+} e^{-i\int_0^\tau \ud \tau' H_{\mathrm{I}}(\tau')}$ can be calculated in time-dependent perturbation theory using the Magnus expansion (reviewed in Appendix~\ref{app:A}) for the Hermitian operator $\mathcal{K}^{(N)}(\tau)=\sum_{n=1}^N K_n(\tau)$. The resulting  $U_{\mathrm{I}}^{(N)}(\tau) =\exp[-i\mathcal{K}^{(N)}(\tau)]$ converges to $U_{\mathrm{I}}(\tau)$ uniformly with growing $N$, and is unitary at all orders~\cite{blanes09}. To first order:
\begin{equation}
K_1(\tau) = \int_0^\tau \ud \tau_1 H_{\mathrm{I}}(\tau_1) = \phi_\tau(E_0 t_f)\splu + \textrm{h.c.}\ ,
\label{eq:K1}
\end{equation}
where
\beq
\phi_\tau(\omega) \equiv  \frac{1}{2}\int_0^{\tau} \ud \tau_1 \dv{\theta}{\tau_1}e^{-i\omega\tau_1}\ .
\label{eq:phi}
\eeq
To systematically go beyond first order we note that the $K_n(\tau)$ are $n$th order nested commutators, and hence  closure of the $su(2)$ Lie algebra guarantees that at all orders $\mathcal{K}^{(N)}(\tau) = \eta^{(N)}(\tau)  \hat{n}^{(N)}(\tau) \cdot\vec{\sigma}$, where $\eta^{(N)}(\tau) >0$, $\hat{n}^{(N)}(\tau) $ is a unit vector, and $\vec{\sigma}=(X,Y,Z)$.
It thus follows that
\beq
U_{\mathrm{I}}^{(N)}(\tau)  = I \cos\eta^{(N)}(\tau)  - i \hat{n}^{(N)}(\tau) \cdot\vec{\sigma} \sin\eta^{(N)}(\tau) \ .
\eeq
We will be concerned primarily with the probability of remaining in the ground state at the final time, denoted $p_{0\leftarrow 0}$. Since $\ket{\psi_{\mathrm{I}}(s)} = U_0^{\dag}(\tau(s)) R_X(\theta(s))\ket{\psi(s)}$, we have $\ket{\psi_{\mathrm{I}}(0)}=\ket{0}$ and $\ket{\psi_{\mathrm{I}}(1)}=-i\ket{0}$. Thus, to $N$th order:
\begin{subequations}
\begin{align}
p^{(N)}_{0\leftarrow 0} &= 1-p^{(N)}_{1\leftarrow 0} = |\bra{0}U^{(N)}(\tau_f) \ket{0}|^2 \\
& = |\cos\eta^{(N)}(\tau_f) - i n_Z^{(n)}(\tau_f)\sin\eta^{(N)}(\tau_f)|^2 \ ,
\end{align}
\end{subequations}
where the states $\ket{0}$ and $\ket{1}$ are the initial ground and excited states,
and where $\tau_f \equiv \tau(1)$.
To first order we find (see App.~\ref{app:A} for the explicit form of $U^{(1)}$):
\begin{align}
\label{eq:mag1}
p^{(1)}_{0\leftarrow 0} = |\bra{0}e^{-i |\phi| X}\ket{0}|^2 = \cos^2(|\phi|) \ , \   \phi \equiv \phi_\tau(E_0 t_f) .
\end{align}
This conceptually elegant result already indicates that quite generally one may expect the ground state probability to oscillate as a function of the anneal time $t_f$, before the adiabatic limit sets in, a conclusion also reached in Ref.~\cite{Brady:2018aa} on the basis of either a large-gap (near-adiabatic limit) or very small gap (stationary phase approximation) assumption. Our analysis applies for arbitrary gaps. 

\begin{figure}[t]
\includegraphics[width=\linewidth]{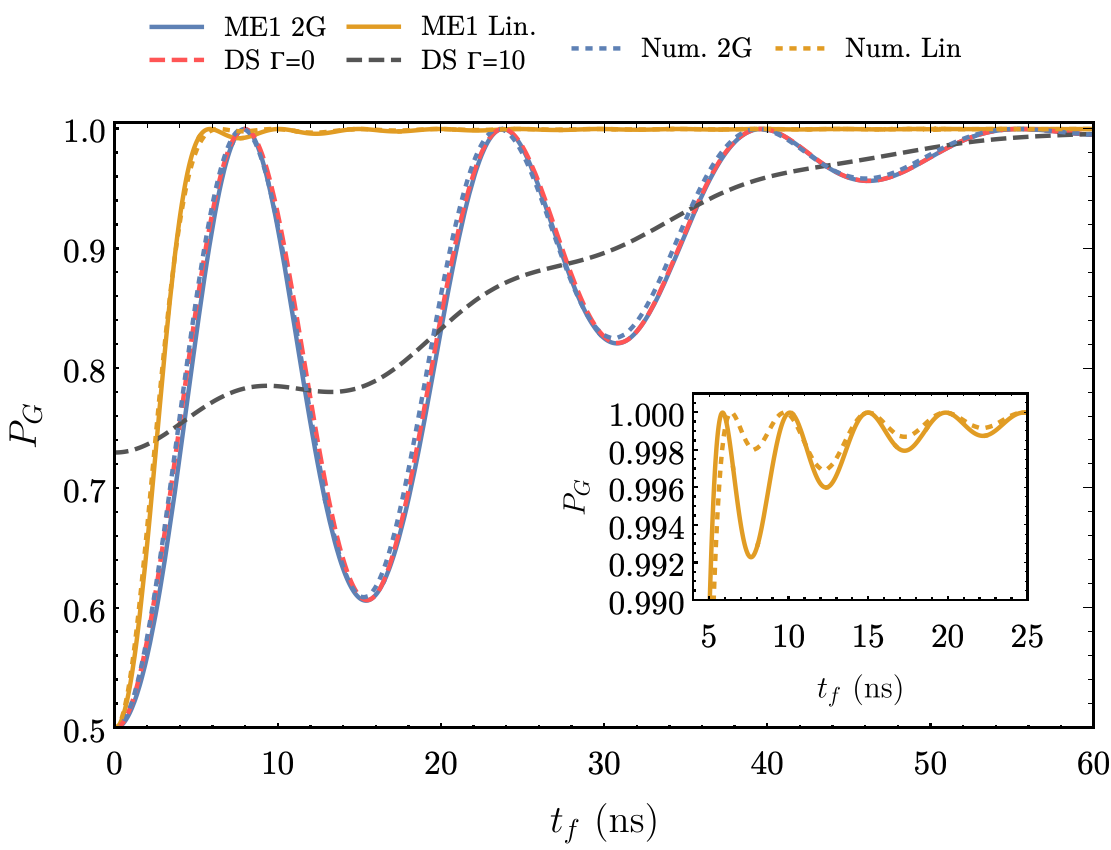}
\caption{(Color online) The numerically exact (dotted) and first order Magnus expansion (solid) ground state probabilities of the linear (orange) and two-step Gaussian progression (blue) at $E_0=0.25\text{ GHz}$. For the two-step Gaussian we set $\alpha=32$ and $\mu=101/800$. Insert: zoomed-in view of the linear schedule results. Here and in other plots we use parameters compatible with quantum annealing using flux qubits~\cite{Dwave,Weber:2017aa,Quintana:2017aa,Novikov:2018aa}. Also shown is the prediction of a simplified double-slit type analysis (dashed, red). Both the latter and the first order Magnus expansion result are in excellent agreement with the numerically exact solution. The effect of strong dephasing in the instantaneous energy eigenbasis is shown as well (dashed, black), obtained using a phenomenological noise model with dephasing parameter $\Gamma$ described in Appendix~\ref{app:D}. In this case the interference oscillations are strongly damped.
}
\label{fig:ts_ph}
\end{figure}

\subsection{LZ problem (linear schedule)}
Let us first consider the simplest annealing schedule, namely a linear interpolation of the type considered in the original LZ problem~\cite{LZ1-1,LZ1-2}: $A(s)=1-s$ and $B(s)=s$. To evaluate Eq.~\eqref{eq:phi} we can change the integration variable to $s$ and approximate $\tau(s)\approx \tau_f s$ in the exponent, yielding $\phi_{\tau_f}(\omega)=\frac{1}{2}\int_{0}^{1} \dd s \frac{1}{s^2 + (1-s)^2}e^{-i\omega \tau_f s}$ for the first-order Magnus expansion. We compare this to the numerically exact solution in Fig.~\ref{fig:ts_ph}, which shows remarkably good agreement. The simplicity of our Magnus expansion approach should be contrasted with the analytical solution for linear schedules in terms of parabolic cylinder functions~\cite{Vitanov:1996uq}. Also notable is that while a quantum interference pattern is visible, the oscillations are very weak and not controllable (see the insert of Fig.~\ref{fig:ts_ph}). This motivates us to introduce new schedules with strong and controllable quantum interference.

\subsection{Strong quantum interference pattern via Gaussian angular progression}
Our goal is to identify a family of annealing schedules that generate strong interference between the paths leading to the final ground state, such that ``giant" oscillations of the ground state probability can be observed. Therefore we now introduce \emph{Gaussian angular progressions}.

Suppose that the angular progression is \emph{two-step Gaussian}, namely, a sum of two Gaussians centered at $\tau_f/2\pm\mu$ (with $\mu < \tau_f/2$):
\beq
\dv{\theta}{\tau} = c\left(
	e^{-[\alpha(\tau-(\tau_f/2+\mu))]^2}
	+e^{-[\alpha(\tau-(\tau_f/2-\mu))]^2}\right)\ .
	\label{eq:sumofG}
\eeq
Note that $\int_0^{\tau_f} d\tau \dv{\theta}{\tau} = \theta(1)-\theta(0)=\frac{\pi}{2}$, which fixes $c$. If we assume that $\alpha \gg 1$ then
we may approximate $\int_0^{\tau_f} $ by $\int_{-\infty}^{\infty}$ (we bound the approximation error in Appendix~\ref{app:B}).
Thus $c=\alpha\sqrt{\pi}/4$ and Eq.~\eqref{eq:phi} yields
$\phi_{\tau_f}(\omega) = \frac{\pi}{4}e^{-i\omega\tau_f/2} e^{-[\omega/(2\alpha)]^2}\cos(\mu\omega)$.
Using Eq.~\eqref{eq:mag1}, to first order the ground state probability is then
\begin{subequations}
\label{eq:2G}
\begin{align}
\label{eq:2Ga}
p^{(1)}_{0\leftarrow 0} &= \cos^2\left[\frac{\pi}{4}e^{-(t_f/t_{\mathrm{ad}})^2}{\cos(\pi t_f/t_{\mathrm{coh}})}\right]\\
\label{eq:2Gb}
t_{\mathrm{ad}} &\equiv 2\alpha/E_0\ , \quad t_{\mathrm{coh}}\equiv \pi/(\mu E_0)\ .
\end{align}
\end{subequations}
The ground state probability thus approaches its adiabatic limit of  $1$ on a timescale of $t_{\mathrm{ad}}$ (set by the Gaussian width), while undergoing damped oscillations with a period of $t_{\mathrm{coh}}$. The oscillations are overdamped when $t_{\mathrm{ad}} < t_{\mathrm{coh}}$. In particular, a single Gaussian ($\mu=0$) can thus not give rise to oscillations.

\begin{figure}[t]
\centering
\subfigure{\includegraphics[width=.9\columnwidth]{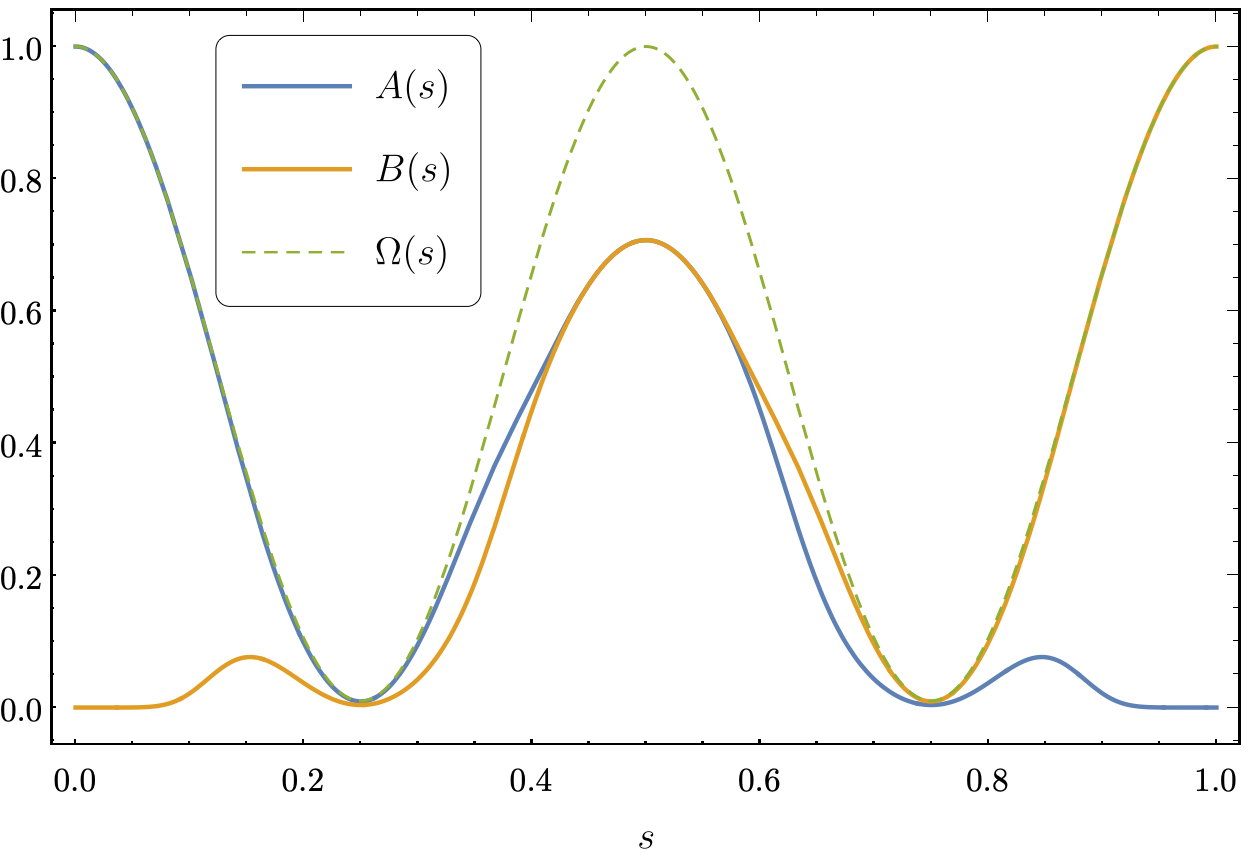}\label{fig:two_step}}
\subfigure{\includegraphics[width=\columnwidth]{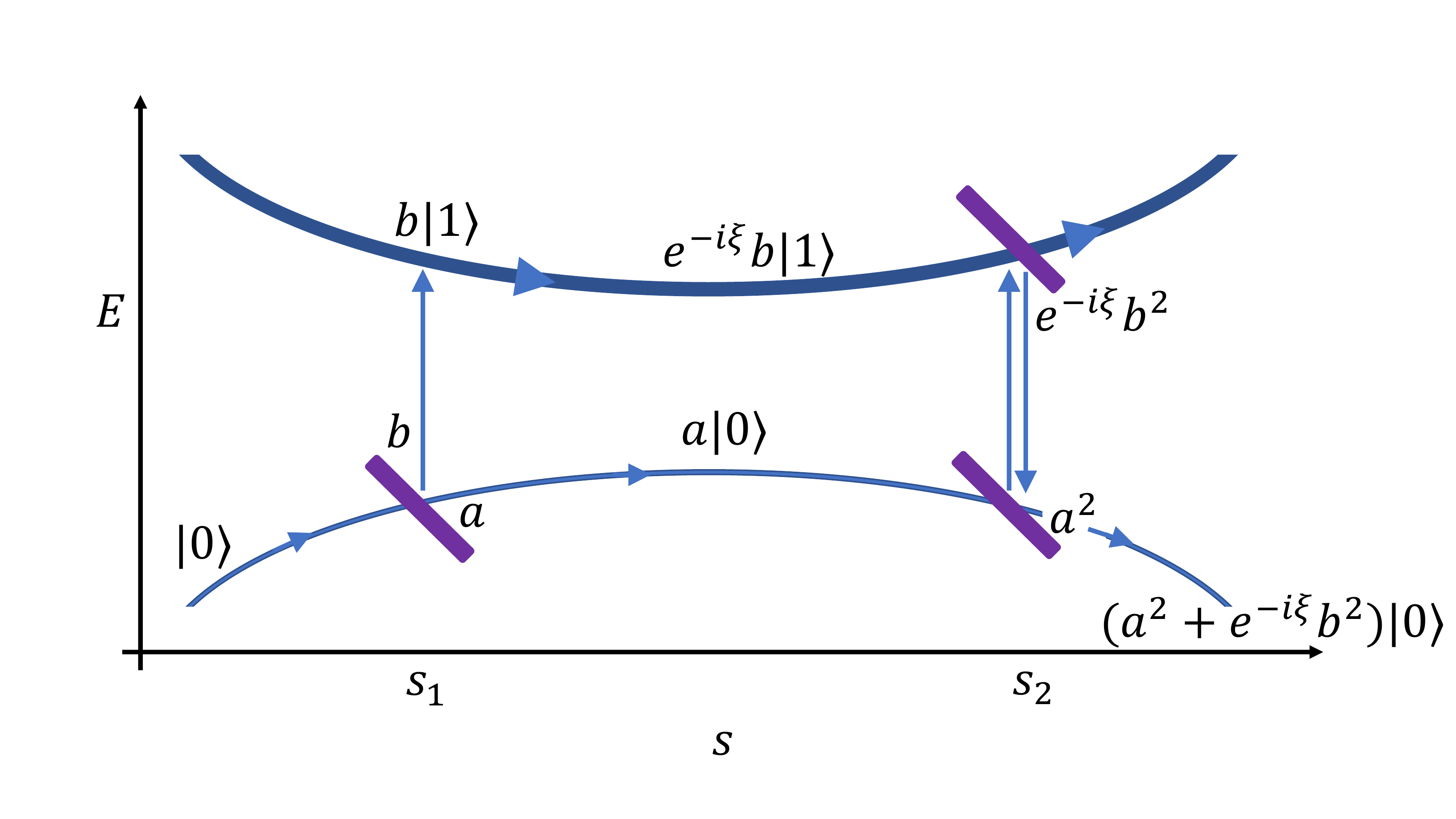}\label{fig:interferometer}}
\caption{(Color online) Top: Example annealing schedules $A(s)$ (blue) and $B(s)$ (orange) for a two-step Gaussian progression with $\alpha=32$ and $\mu=101/800$, subject to the dimensionless gap $\Omega(s)=0.99\cos^2(2\pi s)+0.01$, which is shown as well (dashed, green).
Bottom: Equivalent interferometer model in the adiabatic interaction picture. The system starts in the ground state $\ket{0}$. At $s_1\approx 0.25$ the first Gaussian splits the amplitude, some of which evolves in the excited state $\ket{1}$, where it acquires a relative phase $\xi\propto t_f$. The second Gaussian at $s_2\approx 0.75$ returns part of the excited state amplitude to the ground state, where it recombines. The total ground state amplitude is $a^2+e^{-i\xi}b^2$. Each Gaussian acts as an unbalanced $(a,b)$ beamsplitter (purple), where $a=\cos(\frac{\pi}{8} e^{-(t_f/t_{\mathrm{ad}})^2})$, $b=-i\sin(\frac{\pi}{8} e^{-(t_f/t_{\mathrm{ad}})^2})$ (see Appendix~\ref{app:C} for details).}
\end{figure}

We plot the ground state probability $p_{G}(t_f) \equiv p_{0\leftarrow 0}$ in Fig.~\ref{fig:ts_ph}, for a two-step Gaussian progression with parameters chosen to represent the underdamped case; the associated annealing schedules are shown in Fig.~\ref{fig:two_step}.
The amplitude of the resulting pre-adiabatic oscillations seen in Fig.~\ref{fig:ts_ph} is, as desired, much larger than that associated with the linear schedule. The accuracy of the first-order Magnus expansion is again striking, especially given its simplicity compared to the analytical solution approaches~\cite{Barnes:2012fk,Barnes:2013aa,Messina:2014aa}. We give a bound on the first-order Magnus expansion approximation error in Appendix~\ref{app:B}.

\subsection{Physical origin of the oscillations}
What is the origin of the oscillations? The answer is an interference effect between the two paths created by the two-step schedule, which enforces a double-slit or an unbalanced Mach-Zender interferometer scenario, with $\pi/4$ beam-splitters: see Fig.~\ref{fig:interferometer}. The first step is a perturbation that generates amplitude in the excited state, while the second step allows for some of this amplitude to recombine with the ground state.
The relative phase between the two paths is $\xi=E_0 t_f \int_{s_-}^{s_+}{\Omega(s')}\dd{s'}$, which results in oscillations.  In Appendix~\ref{app:C} we derive this result via a simple interferometer-type model that predicts the curve marked DS $\Gamma=0$ in Fig.~\ref{fig:ts_ph}, which is in excellent agreement with the numerically exact result. 

A natural question is whether the observation of interference oscillations as a function of $t_f$ implies the existence of quantum coherence in the computational basis at $t_f$. We give a formal proof that the answer is affirmative in Appendix~\ref{app:D}. An illustration is given in Fig.~\ref{fig:ts_ph}, for the case of dephasing in the instantaneous energy eigenbasis,
which is equivalent to performing a measurement in this basis between the two Gaussian steps. The final ground state probability is then the sum of \emph{classical} conditional probabilities through each beam-splitter, and as expected, the oscillations disappear.


\subsection{Role of the angular progression}
We emphasize that the angular progression
\beq \label{eq:dthdt}
\odf{\theta}{\tau(s)} =  \frac{B'(s)A(s)-A'(s)B(s)}{\Omega(s)^3}\ ,
\eeq
is the sole quantity needed to determine the ground state probability, per Eqs.~\eqref{eq:phi} and ~\eqref{eq:mag1}. In particular, per Eq.~\eqref{eq:dthdt}, any transformation of $A(s)$, $B(s)$ and $\Omega(s)$ that leaves $\dv{\theta}{\tau}$ invariant will not affect $P_G$ in the closed-system setting. 

Note, furthermore, that specifying the angular progression does not uniquely determine the annealing schedules $A(s)$ and $B(s)$. This is advantageous for practical purposes, since such schedules are typically implemented via arbitrary waveform generators (AWGs) with bandwidth constraints that can be incorporated into the schedule design process. To determine these schedules we need to specify the dimensionless gap $\Omega(s)$ and the angular progression $\dv{\theta}{\tau}$. We can determine $\tau(s)$ by solving the differential equation $\dv{\tau}{s} = \Omega(s)$ subject to the boundary condition $\tau(0) = 0$. Then 
$\theta(s)$ can be determined by solving the differential equation
\beq
\dv{\theta}{s} = \Omega(s)\eval{\dv{\theta}{\tau}}_{\tau=\tau(s)}\ ,
\eeq
subject to appropriate boundary conditions. Together, $\Omega(s)$ and $\theta(s)$ determine the annealing schedules $A(s)$ and $B(s)$ via Eq.~\eqref{eq:2}. In the two-step Gaussian case this means integrating Eq.~\eqref{eq:sumofG}, which, for a constant gap, yields $\theta(s)$ as a sum of $\erf$ functions. 

A particularly interesting example of a dimensionless gap schedule is one that represents the presence of two avoided level crossings, a significant feature of the glued trees problem~\cite{Somma:2012kx}. An example is shown in Fig.~\ref{fig:two_step}, representing an example of the procedure outlined above for numerical determination of the schedule. It is clear from Eq.~\eqref{eq:dthdt} that the main contribution to the angular progression is the near-vanishing of the gap. In contrast, when $\Omega(s)$ is constant, the main contribution to the angular progression is the suddenness of the schedule, i.e., a large $A'(s)$ or $B'(s)$.


\section{Open system analysis and results}
\label{sec:open}
While a phenomenological model of dephasing in the instantaneous energy eigenbasis already shows clearly how the interference pattern disappears under decoherence (Fig.~\ref{fig:ts_ph} and Appendix~\ref{app:D}), this is not a realistic model of decoherence. We thus examine the effect of coupling the TLS to a thermal environment that corresponds more closely to experiments, e.g., with superconducting flux qubits.

We consider a dephasing model wherein the total system-bath Hamiltonian is $H = H_S(t) + H_B + gY\otimes B$, where $B$ is the dimensionless bath operator in the system-bath interaction, $H_S(t)$ is given in Eq.~\eqref{eq:H_S-new},
$H_B$ is the bath Hamiltonian, and $g$ is the coupling strength with units of energy. We assume a separable initial state $\rho_S(0) \otimes \rho_B$, with $\rho_B=\exp(-\beta H_B)/\mathrm{Z}$ the Gibbs state of the bath at inverse temperature $\beta$ and partition function $\mathrm{Z}=\Tr[\exp(-\beta H_B)]$. 
We transform to the interaction picture with respect to $H_B$, so that $H \mapsto \tilde{H}(t) = H_S(t) + g Y \otimes \tilde{B}(t)$, with $\tilde{B}(t) = U_B^\dagger(t) B U_B(t)$, and $U_B(t) = e^{-it H_B}$. The same series of transformations as those leading to Eq.~\eqref{eq:psi-ad} can be summarized as: $Y \otimes \tilde{B}(t) \mapsto t_f Y \otimes \tilde{B}(s) \mapsto t_f R_X(\theta) Y R_X(-\theta)\otimes \tilde{B}(s) =  t_f[\cos(\theta)Y + \sin(\theta)Z]\otimes \tilde{B}(s)$. After the final transformation to the $H_0$-interaction picture, the total Hamiltonian replacing $H_{\mathrm{I}}(\tau)$ in Eq.~\eqref{eq:intp} becomes
\begin{equation}
\label{eq:interaction-h}
	H_{\mathrm{tot}}(s) = \frac{1}{2}\dot{\theta}(s){X}_{\mathrm{I}}(s) + g t_f\vec{\mu}(s)\cdot\vec{\sigma}\otimes\tilde{B}(s)\ ,
\end{equation}
where 
$\vec{\mu} = (\sin\phi\cos\theta,\cos\phi\cos\theta,\sin\theta)$ is a unit vector in polar coordinates, with $\phi(s) \equiv -E_0 t_f \tau(s)$, and henceforth the dot denotes $\odf{}{s}$.

\subsection{Redfield master equation in the adiabatic interaction picture}
The time-convolutionless (TCL) expansion~\cite{Breuer:book} provides a convenient and systematic way to derive master equations (MEs) without requiring an adiabatic or Markovian approximation. With the detailed derivation given in Appendix~\ref{app:E}, the 2nd order TCL (TCL2) ME in the adiabatic-frame can be written as:
\begin{align}
	\dot{\rho}_S(s) &= -i\comm{H_{\mathrm{I}}(s)}{{\rho}_S(s)} \notag \\
	&\qquad -(g t_f)^2 \comm{\vec{{\mu}}(s)\cdot\vec{\sigma}}{\Lambda(s){\rho}_S(s)} + \textrm{h.c.}\ ,
	\label{eq:adiabatic_tcl2}
\end{align}
where
\beq
\Lambda(s) = \int_0^s \dd{s'} C(s,s')U_{\mathrm{I}}(s,s')\vec{{\mu}}(s')U^\dagger_{\mathrm{I}}(s,s')\cdot \vec{\sigma} \ ,
\eeq
and $C(s,s') = \Tr[\tilde{B}(s)\tilde{B}\pqty{s'}\rho_B] = C^*(s',s)$
is the bath correlation function. We assume that the bath is Ohmic  with spectral density
$J\pqty{\omega} = \eta \omega e^{-{\omega}/{\omega_c}}$.
To ensure the validity of the TCL2 approximation---which is also known as the Redfield ME---we derive a general error bound in Appendix~\ref{app:F}, and apply this bound to the Ohmic case. We find the condition $t_f  \ll \frac{\beta}{g^2 \eta}$, which is always satisfied in our simulations.

\subsection{Rotating Wave Approximation (RWA)}
In general, the Redfield ME~\eqref{eq:adiabatic_tcl2} does not generate a completely positive map, which can result in non-sensical results such as negative probabilities~\cite{Gaspard:1999aa,Whitney:2008aa}. Although this is not necessary for complete positivity~\cite{Majenz:2013qw}, a further rotating wave approximation (RWA) is usually performed. The resulting Lindblad-type ME also lends itself to a simpler physical interpretation. As detailed in Appendix~\ref{app:G}, this leads to
\begin{align}
		\dot{\rho}_S &= -i\comm{\frac{1}{2}\dot{\theta}{X}_{\mathrm{I}}+H_{\mathrm{LS}}}{{\rho}_S} \notag \\
		\label{eq:TCL2_RWA}
		&\qquad -g^2 t_f\gamma_d \big(\rho_{ba}\dyad{b}{a}+\rho_{ab}\dyad{a}{b}\big)  \\
		&\qquad + g^2 t_f \gamma_t \pqty{\rho_{aa}-e^{-\beta\Delta}\rho_{bb}}\big(\dyad{b}{b}-\dyad{a}{a}\big) \ , \notag
\end{align}
where $\rho_{ab}=\matrixel{a}{\rho_S}{b}$, all quantities except $g$, $t_f$ and $\beta$ are $s$-dependent, and the effective dephasing and thermalization rates $\gamma_d$ and $\gamma_t$, respectively, and the basis $\{\ket{a},\ket{b}\}$, are given by
\bes
\begin{align}
\label{eq:22a}
\ket{a(s)} &= U_{\mathrm{I}}(s)\ket{\epsilon_-(s)},\ \ket{b(s)} = U_{\mathrm{I}}(s)\ket{\epsilon_+(s)}\\
	\label{eq:22b}
	\gamma_d(s) &= \frac{1}{2} \gamma_t(s)\big(1+e^{-\beta \Delta(s)}\big), \ \gamma_t(s) =\gamma (\Delta(s))\ .
\end{align}
\ees
Here $\ket{\epsilon_\pm(s)}=U_0^\dagger(s)\ket{\pm}$ are the instantaneous eigenvectors of $H_{\mathrm{I}}(s)$.
The Lamb shift is:
\begin{equation}
	H_{\mathrm{LS}}(s) = g^2 t_f(S\pqty{\Delta(s)}\dyad{b}+S\pqty{-\Delta(s)}\dyad{a})\ .
	\label{eq:H_LS}
\end{equation}
The functions $\gamma\pqty{\omega}/2$ and $S\pqty{\omega}$ are the real and imaginary parts of the one-sided Fourier transform of the bath correlation function, and are implicitly $\beta$-dependent (see Appendix~\ref{app:G}, where we also discuss the validity conditions for the RWA).

\subsection{Numerical results}
\begin{figure}[h]
\includegraphics[width=\columnwidth]{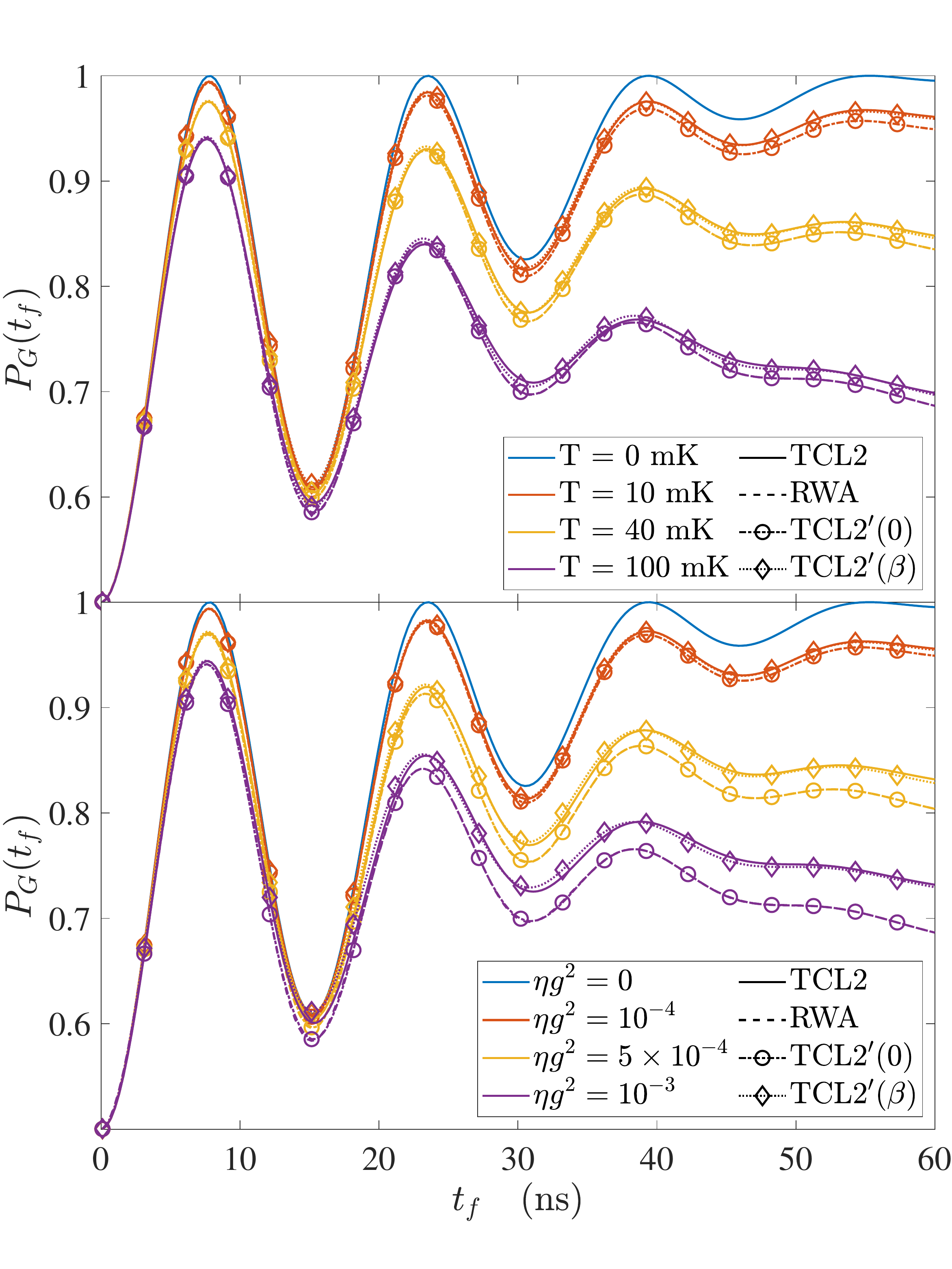}
\caption{(Color online) Ground state probability as a function of total annealing time in the open system setting. Shown are the numerical results of the TCL2 master equation without the RWA [Eq.~\eqref{eq:adiabatic_tcl2}, Redfield] and with the RWA [Eq.~\eqref{eq:TCL2_RWA}, Lindblad], and the semi-empirical Eq.~\eqref{eq:empirical_1}. The bath is Ohmic with a cutoff frequency $\omega_c = 4$ GHZ. Top: $\eta g^2 = 2\times10^{-4}$ for a range of temperatures.  
Bottom: $T=20$ mK for a range of coupling values. $\mathrm{TCL2'}(0)$ is the case $P_E(0)$, and is an excellent agreement with the RWA results. $\mathrm{TCL2'}(\beta)$ is the case $P_E(1/T^*)$ with fitted $T^*$ values. From top to bottom: (a) $T^*=\{13.68, 44.06, 104.50\}$mK and (b) $T^*=\{23.72, 24.22, 22.95\}$mK. Parameter values were chosen to be consistent with quantum annealing using flux qubits and the necessary condition $t_f  \ll \frac{\beta}{g^2 \eta}$.}

\label{fig:redfield_rwa}
\end{figure}

The numerical solutions of Eqs.~\eqref{eq:adiabatic_tcl2} and ~\eqref{eq:TCL2_RWA} are shown in Fig.~\ref{fig:redfield_rwa} for the two-step Gaussian schedule with parameters as in Fig. \ref{fig:ts_ph} and for the gap schedule plotted in Fig.~\ref{fig:two_step}. The main message conveyed by this figure is that oscillations are visible over a wide range (an order of magnitude) of temperatures and system-bath coupling strengths. We also note that for these parameter values the Redfield ME produces physically valid solutions, despite the concerns about complete positivity mentioned above. The Redfield ME results in consistently higher ground state probabilities than the RWA.

These numerical results are accurately reproduced in terms of a simple semi-empirical formula, also shown in Fig.~\ref{fig:redfield_rwa}, and 
derived in Appendix~\ref{app:H}: 
\begin{equation}
\label{eq:empirical_1}
	P'_G(t_f) = \bigg(P_G(t_f)-\frac{1}{2}\bigg)e^{-\bar{\gamma}_d t_f} + P_E(\beta)
\end{equation}
where $P'_G\pqty{t_f}$ and $P_G\pqty{t_f}$ denote the open and closed system success probabilities, respectively, where
\begin{equation}
	\bar{\gamma}_d = g^2 \int_0^1\dd{s'} \gamma_d(s')
	\label{eq:avegamma}
\end{equation}
is the average thermalization rate, and where
\begin{equation}
	P_E(\beta) \equiv \frac{e^{\beta E_0/2}}{Z}\ , \ Z=2\cosh(\beta E_0/2)
\end{equation}
is the ground state probability in the adiabatic limit, given by the thermal equilibrium value associated with $H_S(1)$ [Eq.~\eqref{eq:H_S-new}]. As seen in Fig.~\ref{fig:redfield_rwa}, the agreement is excellent with both the RWA result when we use $P_E(0)=1/2$ (the infinite temperature limit), and with the TCL2 results when we use $P_E(\beta)$ and fit $\beta$; we find that the fitted $\beta$ is consistently slightly lower than the actual $\beta$ values used in our simulations.


\section{Discussion and Conclusions}
\label{sec:conc}
We have proposed a double-slit approach to quantum annealing experiments, exhibiting ``giant"  interference patterns, motivated by the role of coherent diabatic evolution in enabling quantum speedups. Our analytical approach based on a simple time-dependent expansion in the adiabatic interaction picture accurately describes the associated dynamics. The experimental observation of such interference oscillations then becomes a clear and easily testable signature of coherence in the instantaneous energy eigenbasis. The test is simple in principle: it involves a quantum annealing protocol that employs the proposed schedules, with a measurement of only the ground state population as a function of the anneal time $t_f$. When the relative phase between the upper and lower paths to the ground state is randomized, the interference effect is weakened.

To explain these results we proposed an effective model that accurately explains the interference oscillations in terms of a few simple parameters. Namely, upon replacing $P_G(t_f)$ in Eq.~\eqref{eq:empirical_1} by $p^{(1)}_{0\leftarrow 0}(t_f)$ as given in Eq.~\eqref{eq:2Ga}, the three timescales $t_{\mathrm{coh}}$, $t_{\mathrm{ad}}$, and $1/\bar{\gamma}_d$ respectively characterize the oscillation period, Gaussian damping due to approach to the adiabatic limit, and exponential damping due to coupling to the thermal bath. We expressed all three timescales in terms of the input physical parameters of the problem [Eqs.~\eqref{eq:2Gb} and~\eqref{eq:avegamma}], and together they completely characterize the oscillations and their damping.

We expect that an experimental test of our ``double-slit" proposal will reveal the predicted interference oscillations for qubits that are sufficiently coherent, such as aluminum-based flux qubits~\cite{Weber:2017aa,Quintana:2017aa,Novikov:2018aa}, Rydberg atoms~\cite{Glaetzle:2017aa,Pichler:2018aa}, or trapped ions~\cite{Gras:2016aa,Zhang:2017aa}. Such an experiment can be viewed as a necessary condition for quantum annealing implementations of algorithms exhibiting a quantum speedup, e.g., the glued trees problem~\cite{Somma:2012kx}, which rely on coherence between energy eigenstates. It appears relevant (if not essential) to use such coherence in order to bypass the common objection that \emph{stoquastic} quantum annealing or adiabatic quantum computing are subject to, which is that they can be efficiently simulated using the quantum Monte Carlo algorithm when restricted to ground-state evolution (with some known exceptions~\cite{Hastings:2013kk,Jarret:16}), due to the absence of a sign problem~\cite{Albash-Lidar:RMP,Crosson:2018aa}. Therefore an experimental observation of the quantum interference pattern predicted here will bolster our confidence in the abilities of coherent quantum annealers to one day deliver a quantum speedup.

\acknowledgments
We are grateful to L. Campos-Venuti, L. Fry-Bouriaux, M. Khezri, J. Mozgunov, and P. Warburton for insightful comments and discussions. 
We used the Julia programming language~\cite{bezanson_julia:_2017} and the DifferentialEquations.jl package~\cite{rackauckas_differentialequations.jl_2017} for some of the numerical calculations reported in this work.
The research is based upon work (partially) supported by the Office of
the Director of National Intelligence (ODNI), Intelligence Advanced
Research Projects Activity (IARPA), via the U.S. Army Research Office
contract W911NF-17-C-0050. The views and conclusions contained herein are
those of the authors and should not be interpreted as necessarily
representing the official policies or endorsements, either expressed or
implied, of the ODNI, IARPA, or the U.S. Government. The U.S. Government
is authorized to reproduce and distribute reprints for Governmental
purposes notwithstanding any copyright annotation thereon.

\appendix

\section{Dyson and Magnus series}
\label{app:A}

We repeatedly use the following elementary identity for $su(2)$ angular momentum operators:
\begin{align}
\exp(-i\varphi J_{x})J_{z}\exp(i\varphi J_{x}) &
=J_{z}\cos\varphi-J_{y}\sin\varphi\ .
\end{align}
Note that the Pauli matrices are related via $J_{i} = \sigma_i/2$, $i\in\{x,y,z\}$.

Let us denote the solution of the adiabatic frame Hamiltonian given in Eq.~\eqref{eq:psi-ad} by $U_{\mathrm{ad}}(\tau)$. The adiabatic interaction picture propagator, 
\begin{subequations}
\begin{align}
\label{eq:U(tau)}
U_{\mathrm{I}}(\tau) &= U_0^\dagger(\tau)U_{\mathrm{ad}}(\tau) \\
&= T_{+} e^{-i\int_0^\tau \ud \tau' \lambda(\tau')X_{\mathrm{I}}(\tau')}\ ,
\end{align}
\end{subequations}
the solution of Eq.~\eqref{eq:intp}, can be computed using the Dyson series expansion:
\begin{equation}
\label{eq:dyson}
\begin{split}
&U_{\mathrm{I}}(\tau) =  I -i \int_0^{\tau}\ud \tau_1 \lambda(\tau_1)X_{\mathrm{I}}(\tau_1) \\
&\,  +(-i)^2 \int_0^{\tau}\ud \tau_1\int_0^{\tau_1}\ud \tau_2
			 \lambda(\tau_1)X_{\mathrm{I}}(\tau_1) \lambda(\tau_2)X_{\mathrm{I}}(\tau_2)
+\ldots
\end{split}
\end{equation}
Note that each term in the Dyson series
contributes to the ground state amplitude if and only if it is an even power, and likewise to the excitation amplitude if and only if it is an odd power. Consequently, the amplitudes calculated from the Dyson series may not be unitary to a desired precision until the terms are calculated to a high enough order. For this reason we prefer the Magnus expansion~\cite{blanes09}, for which
\beq
U_{\mathrm{I}}(\tau) = \lim_{N\to\infty}\exp[-i\mathcal{K}^{(N)}(\tau)]\ , \quad \mathcal{K}^{(N)}(\tau) = \sum_{n=1}^N K_n(\tau)\ .
\eeq
The first few terms are given by
\bes
\begin{align}
K_1(\tau) &= \int_0^\tau \ud t_1 \lambda(\tau_1) X_{\mathrm{I}}(\tau_1)\\
K_2(\tau) &= -\frac{i}{2} \int_0^\tau \ud \tau_1 \int_0^{\tau_1}\ud \tau_2
	\lambda(\tau_1)\lambda(\tau_2) \comm{X_{\mathrm{I}}(\tau_1)}{X_{\mathrm{I}}(\tau_2)}  \ .
\end{align}
\ees

Using $U_{\mathrm{I}}^{(N)}(\tau) =\exp[-i\mathcal{K}^{(N)}(\tau)]$ and Eq.~\eqref{eq:K1} we thus find
\begin{subequations}
\label{eq:U(1)}
\begin{align}
\label{eq:U(1)a}
U_{\mathrm{I}}^{(1)}(\tau) &= \exp\left(-i [\phi\splu + \textrm{h.c.}]\right) \\
\label{eq:U(1)b}
&=
\begin{pmatrix}
\cos(\abs{\phi}) & -i \sin(\abs{\phi})e^{i\varphi} \\
-i \sin(\abs{\phi})e^{-i\varphi} & \cos(\abs{\phi})
\end{pmatrix} \\
\label{eq:U(1)c}
&= e^{i\varphi Z/2}M_{|\phi|} e^{-i\varphi Z/2}  \\
&  M_{|\phi|}\equiv e^{-i |\phi| X} = \cos(|\phi|)I-i \sin(|\phi|)X\ .
\label{eq:M_phi}
\end{align}
\end{subequations}
where we wrote $\phi$ as a shorthand for $\phi_\tau(E_0 t_f)$, and where $\varphi = \arg(\phi)$. This directly results in Eq.~\eqref{eq:mag1}.

To compute the second order Magnus term we use $X_{\mathrm{I}}(\tau) = e^{-i E_0 t_f \tau}\splu + \textrm{h.c.}$ for the commutation relation

\begin{equation}
\comm{X_{\mathrm{I}}(t_1)}{X_{\mathrm{I}}(t_2)}  =
2i\sin[E_0 t_f (\tau_2 - \tau_1)]Z\ ,
\end{equation}
so that
\begin{equation}
K_2(\tau) = \int_0^{\tau} \ud \tau_1 \int_0^{\tau_1}\ud \tau_2
	\lambda(\tau_1)\lambda(\tau_2) \sin[E_0 t_f(\tau_1 - \tau_2)]Z\ .
	\label{eq:K2}
\end{equation}


\section{Error Analysis of Gaussian Angular Progression Schedules}
\label{app:B}

\subsection{Extension into full Fourier integrals}
We discuss the general Gaussian angular progression
\begin{equation}
\dv{\theta}{\tau} = \psi \frac{\alpha}{\sqrt{\pi}} e^{-\alpha^2(\tau-\mu)^2}
\end{equation}
where $\psi$ is the Bloch sphere rotation angle. In the main text we assumed that we can perform a full Fourier transform (i.e., integration limits extended to $\pm\infty$) to find
\begin{equation}
\phi = \frac{\psi}{2}  e^{-i\omega \mu} e^{-(t_f/t_{\textrm{ad}})^2} 
\end{equation}
and thus arrive at the first order Magnus term
\begin{equation}
K_1 = \frac{\psi}{2} e^{-(t_f/t_{\textrm{ad}})^2}\qty(e^{-i\omega\mu}S_{+} + e^{i\omega\mu}S_{-})\ .
\label{eq:K1-psi}
\end{equation}
We now show that the assumption of a full Fourier transform results in an exponentially small error in $\alpha \tau^*$ where 
\beq
\tau^*=\min\{\mu,\tau_f-\mu\}\ .
\eeq

Let $I$ be the finite time integral
\begin{equation}
I = \int_{0}^{\tau_f}\dd\tau e^{-i\tau \omega} \sqrt{\frac{\alpha^2}{\pi}} e^{-\alpha^2(\tau-\mu)^2}
\end{equation}
with $0<\mu<\tau_f$, and let $F$ be the full Fourier integral
\begin{equation}
F =  \int_{-\infty}^{\infty}\dd\tau e^{-i\tau \omega} \sqrt{\frac{\alpha^2}{\pi}} e^{-\alpha^2(\tau-\mu)^2}
\end{equation}
and define the error
\begin{equation}
\epsilon= \abs{F - I}\ ,
\end{equation}
where
\begin{equation}
F-I = \sqrt{\frac{\alpha^2}{\pi}}
\qty{\int_{-\infty}^{0}+ \int_{\tau_f}^{\infty}}\dd\tau e^{-i\tau\omega} e^{-\alpha^2(\tau-\mu)^2}\ .
\end{equation}
Thus, in terms of the standard normal cumulative density function $\Phi_G(x)=\frac{1}{2}\qty(1+\erf[ x/\sqrt{2}])$,
\bes
\begin{align}
\epsilon 
&\leq
  \sqrt{\frac{\alpha^2}{\pi}}
\qty{\int_{-\infty}^{0}+ \int_{\tau_f}^{\infty}}\dd\tau \abs{ e^{-i\tau\omega} e^{-\alpha^2(\tau-\mu)^2}} \\
&=
\sqrt{\frac{\alpha^2}{\pi}}
\qty{\int_{-\infty}^{0}+ \int_{\tau_f}^{\infty}}\dd\tau e^{-\alpha^2(\tau-\mu)^2}
 \\
&=
\Phi_G \qty( -\sqrt{2} \alpha\mu ) + 1- \Phi_G \qty( \sqrt{2}\alpha(\tau_f -\mu) )\\
&=
 \frac{1}{2}\qty[\erfc(u_1)+ \erfc(u_2)]\ ,
\end{align}
\ees
where $\erfc(x)=1-\erf(x)$ is the complementary error function, and we have set $u_1\equiv \alpha\mu$, and $u_2 \equiv \alpha(\tau_f-\mu)$. The complementary error function is known to have exponentially small bounds \cite{chianiNewExponentialBounds03}. We can quickly derive an even tighter bound by writing the error in terms of the Faddeeva function $w(z)$
\begin{equation}
\epsilon \leq \frac{1}{2}\pqty{e^{-u_1^2} w(i u_1) + e^{-u_2^2} w(i u_2)}
\end{equation}
where
\begin{equation}\label{eq:faddeeva}
\begin{split}
w(z) &= e^{-z^2} \erfc(-i z)\\
&= e^{-z^2}\qty(1+\frac{2 i}{\sqrt{\pi}}\int_0^{z} e^{t^2}\dd t)\ ,
\end{split}
\end{equation}
which is real and positive for imaginary $z$. If $\Im z > 0$, the Faddeeva function has the integral representation \cite[Eq.~7.7.2]{nist-math}
\begin{equation}
w(z) = \frac{i}{\pi} \int_{-\infty}^{\infty}\frac{e^{-t^2}}{z-t}\dd t
\end{equation}
from which we note
\begin{equation}
\abs{w(z)} \leq \frac{1}{\pi} \int_{-\infty}^{\infty}\frac{e^{-t^2}}{\abs{z-t}}\dd t ,
\end{equation}
and as $1/\abs{z-t} \leq 1/\Im z$,
\begin{equation}
\abs{w(z) }\leq \frac{1}{\sqrt{\pi}} \frac{1}{\Im z}\ .
\end{equation}
With this bound on the Faddeeva function, it is straightforward to obtain the error bound
\begin{equation}
\epsilon \leq \frac{1}{2\sqrt{\pi}}\qty(\frac{e^{-u_1^2}}{u_1} + \frac{e^{-u_2^2}}{u_2})\leq \frac{1}{\sqrt{\pi}} 
\frac{e^{-(\alpha\tau^*)^2}}{\alpha\tau^*}\ .
\end{equation}
If $\alpha\tau^* > \sqrt{\pi}\approx 1.77$, then $\epsilon\leq 0.014$. If $\alpha\tau^* > 2\sqrt{\pi}\approx 3.5$, then with  $\epsilon \leq e^{-4\pi}/2\pi$, the percent error in assuming a full Fourier transform is under $0.6$ parts per million. Extending the limits of integration will not result in an appreciable error if $\alpha\tau^*\gtrsim 2$. In other words, if $\alpha^2=1/2\sigma^2$, where $\sigma^2$ is the variance of the normal distribution,  the interval $[0,\tau_f]$ should contain the confidence interval of at least $2\sqrt{2}\sigma\approx 2.8\sigma$.


\subsection{Second Order Term of the Magnus Expansion}

Evaluating the second order of the Magnus expansion [Eq.~\eqref{eq:K2}] yields:
\begin{equation}
\begin{split}
K_2 = \frac{\psi^2}{4} \frac{\alpha^2}{ \pi} \int_0^{\tau_f}\ud\tau_2\int_0^{\tau_2}\ud\tau_1&
e^{-\alpha^2(\tau_1-\mu)^2}e^{-\alpha^2(\tau_2-\mu)^2} \\
& \quad \times
 \sin[\omega(\tau_2-\tau_1)] Z\ .
\end{split}
\end{equation}
The integral is antisymmetric under the exchange of $\tau_1$ and $\tau_2$ due to the sine, so we can extend the time-ordered integrals into the whole square domain as
\begin{equation}
\begin{split}
K_2= \frac{\psi^2}{8} \frac{\alpha^2}{\pi} 
\int_0^{\tau_f}\ud\tau_2
\int_0^{\tau_f}\ud\tau_1
&e^{-\alpha^2(\tau_1-\mu)^2}e^{-\alpha^2(\tau_2-\mu)^2}\\
&\quad\times\sin(\omega \abs{\tau_2-\tau_1}) Z\ .
\end{split}
\end{equation}

For large $\alpha$ we can let $A$ extend over the entire plane. 
The error bound due to extending the integration limits can be found straightforwardly: the square $[0,\tau_f]\times[0,\tau_f]$ contains the circle $C$ centered at $(\mu,\mu)$ with radius $\tau^{*}$, so the region $\mathbb{R}^2-C$ contains a probability mass of $e^{-(\alpha\tau^{*})^2}$ (from a 2D Gaussian distribution). Since $\abs{\sin(\omega \abs{\tau_2-\tau_1})} < 1$, the error in extending the region of integration is therefore bounded by
\begin{equation}
\epsilon_2\leq e^{-(\alpha\tau^{*})^2}\ .
\end{equation}
Thus, up to an error of $\epsilon_2$, we may write
\beq
K_2 = \frac{\psi^2}{8} \ev{\sin(\omega \abs{T_2-T_1}) } Z ,
\eeq
where $T_2$ and $T_1$ are independent Gaussian random variables with variance $1/(2\alpha^2)$. We can perform a change of variables into $T_{+}=T_2+T_1$ and $T_{-} = T_2 - T_1$, which are independent random variables  with sum and difference means $2\mu$ and $0$ respectively, and both with variance $1/\alpha^2$. Finally, the random variable $\abs{T_{-}}$ is known to be distributed according to the folded-normal distribution centered at $0$ (i.e. the half-normal distribution). Thus, the expectation value is precisely the imaginary part of the folded-normal characteristic function  \cite{tsagrisFoldedNormalDistribution14}:
\begin{align}
g_{t_{-}}(\omega) 
&=e^{-\omega^2/2\alpha^2}\times \\
&\ \big[
		( 1-\Phi_G(i \omega/\sqrt{2}\alpha ) ) +( 1-\Phi_G(i \omega/\sqrt{2}\alpha ) ) \big]\ , \notag
\end{align}
where the parent normal distribution has a mean $\mu_{-}=0$. In this case, the characteristic function simplifies to the Faddeeva function
\beq
g_{t_{-}} = e^{-2 r^2}\qty(\erfc(i\sqrt{2} r)) = w\qty(-\sqrt{2} r)\ ,
\eeq
where again $r= \omega/2\alpha$. From  Eq.~\eqref{eq:faddeeva}, we see that
\beq
\Im g_{t_{-}} = -\frac{2}{\sqrt{\pi}} D\qty(\sqrt{2} r)
\eeq
where $D(z)$ is the Dawson function
\begin{equation}
D(z) = e^{-z^2} \int_0^{z} e^{x^2}\dd x\ .
\end{equation}
Thus, as $\ev{\sin(\omega\abs{T_{-}})} = \Im(g_{t_{-}}(\omega))$, the second order term in the Magnus propagator is
\begin{equation}
\label{eq:K2-final}
K_2 = - \frac{\psi^2}{4\sqrt{\pi}} D\qty(\sqrt{2} r) Z\ .
\end{equation}


\subsection{Magnus Expansion Convergence and Error Bounds}

Let $S \geq \norm{K_1}$ be a bound on the operator norm of the first order term in the Magnus expansion. A sufficient condition for the convergence of the Magnus expansion is that \cite{blanesMagnusFerExpansions1998}
\begin{equation}
S \leq \xi = 1.08686870\ .
\end{equation}
With a Gaussian angular progression as given in Eq.~\eqref{eq:K1-psi}, and noting that $\norm{e^{-i\omega\mu}S_{+} + e^{i\omega\mu}S_{-}}\leq 1$, it is then sufficient that
\begin{equation}
S = \frac{\psi}{2} e^{-(t_f/t_{\textrm{ad}})^2} < \xi
\end{equation}
for the Magnus expansion to be convergent. This means that $\psi \leq 2\xi$, and in particular the physically relevant range $\psi\in[0,\pi/2]$ ($\pi/2$ represents a balanced beam-splitter, and $\psi>\pi/2$ is equivalent to $\pi-\psi$) is within the convergence radius.

If $\mathcal{K}^{(n)}$ is the $n$th order truncation of the Magnus expansion, the error in the truncation is given by
\begin{equation}
\epsilon_{\text{ME}(n)}\equiv \norm{\mathcal{K} - \mathcal{K}^{(n)}} \leq \sum_{m=n+1}^{\infty} S^{m} b_{m}
\end{equation}
where $\{b_m\}$ is a sequence defined in Ref.~\cite{blanesMagnusFerExpansions1998} via various recurrence relations. For $\psi = \pi/2$ and for $\omega/(2\alpha)=0,0.5$ and $1.0$, the corresponding second order truncation errors are $0.25, 0.1$, and $0.008$.


\section{Double-slit interpretation}
\label{app:C}

Having derived the adiabatic frame Hamiltonian given in Eq.~\eqref{eq:psi-ad}
\beq
H_{\mathrm{ad}}(\tau) =\frac{1}{2}\left(\dv{\theta}{\tau}X - E_0 t_f Z\right) \ ,
\label{eq:C1}
\eeq
we see that the angular progression $\dv{\theta}{\tau}$ of an annealing schedule is the perturbation that causes transitions between the two levels of the system. While this perturbation is steady and small in the case of a linear schedule, Gaussian schedules in which the perturbation is localized suggest an appealing physical picture similar to a double-slit or interferometer model. 


\subsection{Single Gaussian step}

Let us first consider a \emph{single} Gaussian step, which Eq.~\eqref{eq:sumofG} reduces to when $\mu=0$, $c=\alpha\sqrt{\pi}/2$. Under the same assumptions as those leading to Eq.~\eqref{eq:2G}, we then find $\phi_{\tau_f}(\omega) = \frac{\pi}{4}e^{-i\omega\tau_f/2} e^{-(t_f/t_{\textrm{ad}})^2}$, with $\omega = E_0 t_f$.
Thus, Eq.~\eqref{eq:U(1)} gives us the first order Magnus expansion propagator in the interaction picture with 
\beq
|\phi| = \frac{\pi}{4} e^{-[E_0 t_f/(2\alpha)]^2}= \frac{\pi}{4} e^{-(t_f/t_{\mathrm{ad}})^2}
\label{eq:phi-pi4}
\eeq 
and $\varphi =E_0 t_f \tau_f/2$. The $X$-rotation matrix in Eq.~\eqref{eq:U(1)c} thus becomes:
\begin{equation}
\label{eq:beam_splitter_matrix-0}
M^{\mathrm{G}}_{\psi}=
\begin{pmatrix}
\cos(\frac{\psi}{2} e^{-(t_f/t_{\mathrm{ad}})^2}) &
-i \sin(\frac{\psi}{2} e^{-(t_f/t_{\mathrm{ad}})^2}) \\
-i \sin(\frac{\psi}{2} e^{-(t_f/t_{\mathrm{ad}})^2}) &
\cos(\frac{\psi}{2} e^{-(t_f/t_{\mathrm{ad}})^2}) 
\end{pmatrix} \ ,
\end{equation}
with the superscript G serving as a reminder that this is the Gaussian step case.
Now let us suppose that the Gaussian profile is narrow: $\alpha \gg E_0 t_f$, or equivalently $t_{\mathrm{ad}} \gg t_f$. 
The perturbation is then sudden relative to the adiabatic timescale, and acts like a beamsplitter in a Mach-Zehnder (MZ) interferometer~\cite{Oliver:2005aa}. In this limit $|\phi| \approx \pi/4$ and Eq.~\eqref{eq:U(1)c} gives 
\begin{align}
\label{eq:beam_splitter_matrix}
U_{\mathrm{I}}^{(1)}(\tau_f) 
&=
e^{i(E_0 t_f \tau_f/2) Z} M^{\mathrm{G}}_{\pi/2}  e^{-i(E_0 t_f \tau_f/2)Z } \\
&\quad M^{\mathrm{G}}_{\pi/2} = \frac{1}{\sqrt{2}}
\begin{pmatrix}
1 & -i \\
-i & 1
\end{pmatrix} \ . \notag
\end{align}
Recall that in the adiabatic interaction picture $\ket{\psi_{\mathrm{I}}(0)}=\ket{0}$. 
Thus, the first phase factor $e^{-i\varphi Z}$ has no effect, and we can picture a process by which the ground state $\ket{0}$ is instantly split into an equal superposition $\frac{1}{\sqrt{2}}(\ket{0}-i\ket{1})$ by the ``Mach-Zender" matrix $M^{\mathrm{G}}_{\pi/2}$.
These two states are then propagated freely by $U^\dagger_0(\tau_f) = e^{i(E_0 t_f \tau_f/2) Z}$, so they accumulate a relative phase of $i e^{i E_0 t_f \tau_f}$. For a single Gaussian, interference due to this phase difference is clearly not picked up via a $Z$ basis measurement.


\subsection{Two Gaussian steps: indirect derivation of the interferometer model in the narrow Gaussian limit} 

If instead we consider a two-step Gaussian schedule [Eq.~\eqref{eq:sumofG}], 
then as we already found before Eq.~\eqref{eq:2G}, $\phi_{\tau_f}(\omega) = \frac{\pi}{4}e^{-i\omega\tau_f/2} e^{-(t_f/t_{\textrm{ad}})^2}\cos(\mu\omega)$, with $\omega = E_0 t_f$. 
Eq.~\eqref{eq:U(1)} now gives us the first order Magnus expansion propagator in the interaction picture with $|\phi| = \frac{\pi}{4} |\cos(\mu E_0 t_f)|e^{-(t_f/t_{\textrm{ad}})^2}$ and again $\varphi =E_0 t_f \tau_f/2$.%
\footnote{Note that without the exponential decay factor $e^{-(t_f/t_{\textrm{ad}})^2}=e^{-(t_f/t_{\mathrm{ad}})^2}$ the oscillations are completely undamped and the adiabatic limit is never reached. Thus it is clear that the \emph{finite} width of the Gaussian steps is solely responsible for the onset of adiabaticity.}

Let us now derive an equivalent MZ interferometer model.
On the one hand, we already know from Eq.~\eqref{eq:mag1} that $p^{(1)}_{0\leftarrow 0}= \cos^2(|\phi|)$, i.e.
\beq
p^{(1)}_{0\leftarrow 0}=\cos^2(\frac{\pi}{4} |\cos(\mu E_0 t_f)|e^{-(t_f/t_{ad})^2})\ .
\label{p-BS}
\eeq 
This function has a quasiperiod (the distance between consecutive maxima) of $\pi/(\mu E_0)$, a minimum of $\cos^2(\pi/4)=1/2$ at $t_f=0$, and a maximum of $1$. 
On the other hand, we may model the two-step narrow ($\alpha \gg E_0 t_f$) Gaussian schedule as two consecutive, localized (at $\tau_f/2\pm\mu$) and non-overlapping ($\alpha \gg 1/\mu$) ``beam-splitter" steps, separated by a dimensionless time interval of $2\mu$. Each beam-splitter is of the form given in Eq.~\eqref{eq:beam_splitter_matrix}, the only difference being that the first acts at $\tau_f/2-\mu$ (preceded by free evolution) and the second acts at $\tau_f/2+\mu$ (followed by free evolution). In between the beam-splitter action there is free evolution of duration $2\mu$. Ignoring the initial and final free evolutions (since the initial and final state we are interested are both $\ket{0}$, which is invariant under $U_0$) we expect to be able to write the propagator as the following ansatz:
\begin{align}
\label{eq:beam_splitter_matrix-2}
& \tilde{U}^{(1)}(\tau_f) =M^{\textrm{G}}_{\psi} U_0(2\mu) M^{\textrm{G}}_{\psi}   
\end{align}
where we left the angle $\psi$ in the beam splitter matrix~\eqref{eq:beam_splitter_matrix-0} unspecified in order to determine it by matching to the properties of 
$p^{(1)}_{0\leftarrow 0}= \cos^2(|\phi|)$. Carrying out the matrix multiplication and computing the expectation value, we find
\beq
\abs{\bra{0}\tilde{U}^{(1)}(\tau_f)\ket{0}}^2 = \abs{\cos^2(\psi/2) - \sin^2(\psi/2) e^{2i\mu E_0 t_f}}^2\ .
\label{eq:interf}
\eeq
In order for this to match Eq.~\eqref{p-BS}, we require a quasiperiod of $\pi/(\mu E_0)$ (which is already the case), a minimum of $1/2$ at $t_f=0$, and a maximum of $1$. The latter two conditions force $\psi=\pi/4$.

Therefore, considering Eq.~\eqref{eq:beam_splitter_matrix-2}, we have shown that the two-step Gaussian model is equivalent (in the large $\alpha$ limit) to a MZ interferometer with two unbalanced beamsplitters, separated by free propagation of duration $2\mu$ (the separation between the two Gaussians).

The double-slit (or MZ interferometer model) is remarkably accurate in terms of predicting the ground state probability. This is shown in Fig.~\ref{fig:ts_ph}, where we compare the numerically exact result and the solution of the simple interferometer model given by Eq.~\eqref{eq:interf}.  
Namely, we use the interference model given in Eq.~\eqref{eq:interf}, with $\psi=\pi/4$. To calculate the interference fringe, the position of each of the two Gaussians is given by $s_{\pm}=(\tau_f/2\pm\mu)/\tau$. The  phase factor $\mu E_0 t_f$, which only holds in the large $\alpha$ limit, is replaced by $E_0 t_f [\tau(s_+)-\tau(s_-)] = E_0 t_f \int_{s_-}^{s_+}\dd{s'}\Omega(s')$, where $\tau(s)$ is the cumulative dimensionless gap [Eq.~\eqref{eq:3}]. The reason for this replacement is given in the following, alternative and more direct derivation of the interferometer model.

\subsection{Two Gaussian steps: direct derivation of the interferometer model}

Given the two-step Gaussian schedule, Eq.~\eqref{eq:sumofG}, 
\beq
\dv{\theta}{\tau} = c\left(
	e^{-[\alpha(\tau-\tau_+)]^2}
	+e^{-[\alpha(\tau-\tau_-)]^2}\right)\ ,
\eeq
where $\tau_\pm = \tau_f/2\pm\mu$,
we can split the unitary generated by the adiabatic frame Hamiltonian, Eq.~\eqref{eq:C1}, into two parts:
\begin{equation}
\label{eq:U_split}
    U_{\mathrm{ad}}(\tau_f,0)=U_{\mathrm{ad}}(\tau_f, \frac{\tau_f}{2})U_{\mathrm{ad}}(\frac{\tau_f}{2}, 0)
\end{equation}
We now wish to apply the Magnus expansion separately to each of the unitaries $U_{\mathrm{ad}}(\frac{\tau_f}{2}, 0)$ and $U_{\mathrm{ad}}(\tau_f, \frac{\tau_f}{2})$. 
Consider $U_{\mathrm{ad}}(\frac{\tau_f}{2}, 0)$. 
Inverting Eq.~\eqref{eq:U(tau)}, the first order Magnus expansion [Eq.~\eqref{eq:U(1)}] gives
\bes
\begin{align}
    U_{\mathrm{ad}}(\frac{\tau_f}{2},0) &= U_0(\frac{\tau_f}{2},0)U_{\mathrm{I}}^{(1)}(\frac{\tau_f}{2},0) \\
    &=U_0(\frac{\tau_f}{2},0) e^{i\varphi Z/2}M_{|\phi|} e^{-i\varphi Z/2} \ ,
\end{align}
\ees
where, using Eq.~\eqref{eq:phi}, now
\begin{equation}
    \phi\equiv \phi_{\tau_f/2,0}(E_0 t_f) = \frac{1}{2} \int_{0}^{{\tau_f}/{2}}\dv{\theta}{\tau_1}e^{-iE_0 t_f\tau_1} \dd{\tau_1}.
\end{equation}
For $\alpha \gg 1$ we may extend the limits of integration over the interval $[0,\tau_f/2]$ to $\pm\infty$ without considering the second Gaussian step:
\bes
\begin{align}
    \phi &\approx \frac{c}{2}\int_{-\infty}^{\infty}e^{-[\alpha (\tau_1-\tau_-)]^2} e^{-iE_0 t_f\tau_1} \dd{\tau_1} \\
    &=\frac{\pi}{8}e^{-i E_0 t_f \tau_-}e^{-(t_f/t_{\mathrm{ad}})^2} ,
    \label{eq:phi-pi8}
\end{align}
\ees
where we used $c=\alpha\sqrt{\pi}/4$ as we found in the derivation of Eq.~\eqref{eq:2G}.
We may thus write the explicit form of the interaction picture unitary as
\bes
\begin{align}
    U_{\mathrm{I}}^{(1)}(\frac{\tau_f}{2},0)&= e^{{i(E_0t_f\tau_-}/{2})Z} M^{\mathrm{G}}_{\pi/4} e^{{-i(E_0t_f\tau_-}/{2})Z} \\
    &= U^\dagger_0(\tau_-, 0) M^{\mathrm{G}}_{\pi/4} U_0(\tau_-, 0) \ ,
\end{align}
\ees
and the adiabatic frame unitary becomes:
\bes
\begin{align}
    U_{\mathrm{ad}}(\frac{\tau_f}{2},0) &= U_0(\frac{\tau_f}{2},0) U_0^\dagger(\tau_-, 0) M^{\mathrm{G}}_{\pi/4} U_0(\tau_-, 0) \\
    &= U_0(\frac{\tau_f}{2},\tau_-) M^{\mathrm{G}}_{\pi/4} U_0(\tau_-, 0)\ .
\end{align}
\ees

Repeating this calculation for the second adiabatic frame unitary $U_{\mathrm{ad}}(\tau_f, \frac{\tau_f}{2})$, we obtain
\begin{equation}
    U_{\mathrm{ad}}(\tau_f, \frac{\tau_f}{2}) = U_0(\tau_f,\tau_+) M^{\mathrm{G}}_{\pi/4} U_0(\tau_+, \frac{\tau_f}{2})\ .
\end{equation}
Thus, Eq.~\eqref{eq:U_split} becomes
\begin{equation}
\label{eq:beam_splitter_unitaries}
    U_{\mathrm{ad}}(\tau_f, 0) = U_0(\tau_f,\tau_+) M^{\mathrm{G}}_{\pi/4} U_0(\tau_+, \tau_-)  M^{\mathrm{G}}_{\pi/4} U_0(\tau_-, 0)\ ,
\end{equation}
which describes an interferometer composed of two unbalanced ($\pi/4$) double beam-splitters, interrupted by free propagation of duration $\tau_+-\tau_-$ (ignoring the initial and final phases).

The phase accumulated between $\ket{0}$ and $\ket{1}$ is solely determined by the free evolution in Eq.~\eqref{eq:beam_splitter_unitaries},
\begin{equation}
    U_0(\tau_+, \tau_-) = e^{i[{E_0 t_f}(\tau_+ -\tau_-)/2]Z}
    \label{eq:U0}
\end{equation}
whose value is given by 
\begin{equation}
\label{eq:accumulated_phase}
    \xi = E_0 t_f (\tau_+ - \tau_-) = E_0 t_f \int_{s_-}^{s_+}{\Omega(s')}\dd{s'} \ ,
\end{equation}
where in the second equality we used Eq.~\eqref{eq:3}.

\section{Interference oscillations in the double-slit experiment imply quantum coherence in the computational basis}
\label{app:D}

Here we prove that coherence in the energy eigenbasis implies, in general, coherence in the computational basis.

Let $H(t)$ denote an arbitrary, time-dependent TLS Hamiltonian, with instantaneous energy eigenbasis $\{\ket{\epsilon_i(t)}\}$. The TLS density matrix can be written in this basis as
\begin{equation}
    \rho(t) = \sum_{ij}\tilde{\rho}_{ij}(t)\dyad{\epsilon_i(t)}{\epsilon_j(t)} \ .
\end{equation}
Let us define ``coherence" with respect to a given basis as the off-diagonal elements of the density matrix in the same basis. We can compute the coherence in the computational basis $\{\ket{0},\ket{1}\}$ via
\begin{equation}
\label{eq:computational_coherence}
    \rho_{01} = \mel{0}{\rho(t)}{1} = \sum_{ij} \mel{0}{\tilde{\rho}_{ij}(t)\epsilon_{ij}(t)}{1} ,
\end{equation}
where $\epsilon_{ij}(t) = \dyad{\epsilon_i(t)}{\epsilon_j(t)}$. The two bases are related via a unitary rotation:
\bes
\begin{align}
    \ket{\epsilon_0(t)} &= \cos{\theta\pqty{t}}\ket{0} + e^{i\phi(t)}\sin{\theta\pqty{t}} \ket{1}\\
    \ket{\epsilon_1(t)} &= \sin{\theta\pqty{t}}\ket{0} - e^{i\phi(t)}\cos{\theta\pqty{t}} \ket{1} \ ,
\end{align}
\ees
so that Eq,~\eqref{eq:computational_coherence} reduces to:
\begin{align}
    \mel{0}{\rho(t)}{1} 
	&=e^{-i\phi}\Big\{(\tilde{\rho}_{00}-\frac{1}{2})\sin(2\theta)-\Re(\tilde{\rho}_{10})\cos(2\theta) \nonumber\\
	&+ i\Im(\tilde{\rho}_{10}) \Big\} \ . \label{eq:D4}
\end{align}
where we used $	\tilde{\rho}_{00}+\tilde{\rho}_{11}=1$ and $\tilde{\rho}_{01} = \tilde{\rho}_{10}^*$.
Equation~\eqref{eq:D4} can be further simplified using
$(\tilde{\rho}_{00}-\frac{1}{2})\sin(2\theta) -\Re(\tilde{\rho}_{10})\cos(2\theta) =
	C(\cos\varphi\sin(2\theta)-\sin\varphi\cos(2\theta))$,
where
\bes
\begin{align}
    &C = \sqrt{(\Re\tilde{\rho}_{10})^2+(\tilde{\rho}_{00}-\frac{1}{2})^2} \\
    &\tan\varphi = \frac{\Re(\tilde{\rho}_{10})}{\tilde{\rho}_{00}-\frac{1}{2}} \ .
\end{align}
\ees
Additionally, by making use of the trigonometric identity $\sin(2\theta - \varphi) = \sin2\theta\cos\varphi - \sin\varphi\cos2\theta$, 
Eq.~\eqref{eq:D4} can be written as
\begin{align}
	\mel{0}{\rho(t)}{1} = e^{-i\phi}(C\sin(2\theta - \varphi) + i\Im\tilde{\rho}_{10}) \ . \label{eq:off_diag_computational}
\end{align}
Since $C\sin(2\theta - \varphi) \in \mathds{R}$, it follows that $\Im(\tilde{\rho}_{10}(t)) \neq 0$ implies $\mel{0}{\rho(t)}{1} \neq 0$. Therefore we next establish that indeed, $\Im(\tilde{\rho}_{10}(t)) \neq 0$ in our double-slit proposal. 

Consider the the ground state just before the first beam-splitter,
\begin{equation}
    \rho(\tau_- -\varepsilon) = \dyad{\epsilon_0}
\end{equation}
with $\varepsilon/(\tau_+-\tau_-)\ll 1$. This state evolves through the double-beam-splitter region [recall Eq.~\eqref{eq:beam_splitter_unitaries}]:
\begin{equation}
    M_{|\phi|}U_0(\tau_+,\tau_-)M_{|\phi|} \ ,
\end{equation}
where $U_0$ is given in Eq.~\eqref{eq:U0} and $M_{|\phi|}$ is given in Eq.~\eqref{eq:M_phi}. 

After passing through the first beam-splitter, the system density matrix in the energy eigenbasis becomes
\begin{equation}
    \rho(\tau_-+\varepsilon) = \mqty(\cos^2(\abs{\phi}) & i\sin(\abs{\phi})\cos(\abs{\phi}) \\ -i\sin(\abs{\phi})\cos(\abs{\phi}) & \sin^2(\abs{\phi})) .
\end{equation}
It is useful to include a simple model of decoherence between energy eigenstates during the time interval $[\tau_-, \tau_+]$, complementary to our master equation treatment. We can do so by introducing a continuous dephasing channel. This damps the phases by the factor $e^{-\Gamma\Delta \tau}$, where $\Delta \tau = \tau_+- \tau_- = 2\mu$, and $\Gamma>0$ is the dephasing rate. Right before the second beam-splitter, the system density matrix is then:
\begin{widetext}
\begin{equation}
        \rho(\tau_+-\varepsilon) = \mqty(\cos^2(\abs{\phi}) & ie^{-\Gamma\Delta \tau}e^{it_fE_0\Delta \tau}\sin(\abs{\phi})\cos(\abs{\phi}) \\ -ie^{-\Gamma\Delta \tau}e^{-it_fE_0\Delta \tau}\sin(\abs{\phi})\cos(\abs{\phi}) & \sin^2(\abs{\phi})) \ .
\end{equation}
After passing through the second beam-splitter, the state becomes $\rho(\tau_+ +\varepsilon)  = M_{|\phi|}\rho(\tau_+-\varepsilon)M^\dagger_{|\phi|}$. We find, after some algebra:
\begin{subequations}
\label{eq:coherence}
\begin{align}
\label{eq:coherence1}
P_G &= \tilde{\rho}_{00} = \sin ^4(|\phi|)+\cos ^4(|\phi| ) -2 e^{-\Gamma  \Delta \tau } \sin ^2(|\phi| ) \cos ^2(|\phi|) \cos (\Delta \tau  {E_0 t_f}) \stackrel{\Gamma\to\infty}{\longrightarrow} \frac{1}{4} [\cos (4 |\phi|)+3]\\
\label{eq:coherenc2}
\tilde{\rho}_{01} &= \frac{1}{2} \sin (2 |\phi| ) \left(e^{-\Gamma  \Delta \tau } [-\sin (\Delta \tau  {E_0 t_f})+i \cos (2 |\phi| ) \cos (\Delta \tau  {E_0 t_f})]+i \cos (2 |\phi| )\right) \stackrel{\Gamma\to\infty}{\longrightarrow} i\frac{1}{4}\sin(4|\phi| ) .
\end{align}
\end{subequations}
\end{widetext}
We now note from Eq.~\eqref{eq:phi-pi8} that $\abs{\phi} = \frac{\pi}{8}e^{-(t_f/t_{\mathrm{ad}})^2}$. Therefore we may conclude that $\Im(\tilde{\rho}_{10}(t_f)) > 0$, and  $\Im(\tilde{\rho}_{10}) \to 0$ only in the adiabatic limit ($t_f \gg t_{\mathrm{ad}}$, which implies $\abs{\phi}\to 0$). Note that Eq.~\eqref{eq:coherence1} generalizes Eq.~\eqref{eq:interf} by including the effect of dephasing in the energy eigenbasis.

It is clear from Eq.~\eqref{eq:coherence} that oscillations in the ground state probability $P_G(t_f)$, which are present for finite $\Gamma$, imply a non-vanishing $\Im(\tilde{\rho}_{10}(t_f))$. Therefore we may conclude that the observation of interference oscillations in our proposed double-slit experiment are also evidence of coherence in the computational basis at $t_f$. For finite $\Gamma$, such coherence vanishes only in the adiabatic limit.


\section{Derivation of the adiabatic-frame TCL2/Redfield master equation}
\label{app:E}

We start from the Hamiltonian given in Eq.~\eqref{eq:interaction-h}, which we write as 
\begin{subequations}
\begin{align}
H_{\mathrm{tot}}(s) &=  H_{\mathrm{I}}(s) + \tilde{H}_{SB}(s)\\
&H_{\mathrm{I}}(s) = \frac{1}{2}\dot{\theta}(s){X}_{\mathrm{I}}(s)\\
&\tilde{H}_{SB}(s) =\kappa \vec{\mu}(s)\cdot\vec{\sigma}\otimes\tilde{B}(s)\ ,
\end{align}
\end{subequations}
where $\kappa \equiv g t_f$.
Our goal is to derive a master equation for the system evolution. It is convenient to do so using the time-convolutionless (TCL) approach~\cite{Breuer:book}. 
To do so we must first perform yet another interaction picture transformation, defined by $H_{\mathrm{I}}(s)$, with the associated unitary $U_{\mathrm{I}}(s,s') = T_+ \exp[-i\int^s_{s'} H_{\mathrm{I}}(s'')ds'']$, where $T_{+}$ denotes forward time-ordering. In this frame the total Hamiltonian $H_{\mathrm{tot}}(s)$ becomes 
\beq
\tilde{H}_{\mathrm{tot}}(s) = \kappa \vec{\tilde{\mu}}(s)\cdot\vec{\sigma}\otimes\tilde{B}(s),\quad \vec{\tilde{\mu}}(s) = U^\dagger_{\mathrm{I}}(s,0)\vec{\mu}(s)U_{\mathrm{I}}(s,0) .
\eeq
We can now calculate the TCL expansion generated by the superoperator
\begin{equation}
	\mathcal{L}(s)\rho = 
	-i  \comm{\tilde{H}_{\mathrm{tot}}(s)}{\rho},
\end{equation}
whereupon 
\beq
\dot{\tilde{\rho}}(s) = \sum_{n=1}^\infty \kappa^{2n} \mathcal{K}_{2n}(s) \tilde{\rho}(s) .
\label{eq:IPME}
\eeq 
The different orders are called TCL2, TCL4, etc. We give details on the convergence criteria of this expansion in Appendix~\ref{app:F}.

To second order the TCL generator is:
\begin{align}
\label{eq:tcl2_generator}
	&\mathcal{K}_2(s) [\tilde{\rho}_S\otimes\rho_B]  \\
	&=- \int_0^s\dd{s'}\Tr_B\comm{\tilde{H}_{\mathrm{tot}}(s)}{\comm{\tilde{H}_{\mathrm{tot}}\pqty{s'}}{\tilde{\rho}_S(s)\otimes\rho_B}} , \notag
\end{align}
where $\rho_B$ is the initial state of the bath, and the joint initial state is assumed to be in the factorized form $\rho_S\otimes \rho_B$. Note that the TCL2 approximation coincides with the Redfield master equation~\cite{Breuer:book}.

Let 
\beq
C(s,s')=\Tr [\tilde{B}(s)\tilde{B}\pqty{s'}\rho_B] = C^*(s',s)
\eeq 
denote the bath correlation function.
By explicitly tracing out the bath, $\mathcal{K}_2(s)$ can be written as
\begin{equation}
	\mathcal{K}_2(s)\tilde{\rho}_S = -\kappa^2 \comm{\vec{\tilde{\mu}}(s)\cdot\vec{\sigma}}{\tilde{\Lambda}(s)\tilde{\rho}_S} + \textrm{h.c.}
\end{equation}
where
\begin{align}
	\tilde{\Lambda}(s) =  \int_0^s \dd{s'} C(s,s') \vec{\tilde{\mu}}(s')\cdot\vec{\sigma} .
\end{align}
After transforming back to the Schr\"odinger frame with respect to $H_{\mathrm{I}}(s)$ we obtain:
\begin{align}
	\dot{\rho}_S(s) &= -i\comm{H_{\mathrm{I}}(s)}{{\rho}_S(s)} \notag \\
	&\qquad -\kappa^2 \comm{\vec{{\mu}}(s)\cdot\vec{\sigma}}{\Lambda(s){\rho}_S(s)} + \textrm{h.c.},
	\label{eq:C9}
\end{align}
where
\beq
\Lambda(s) = \int_0^s \dd{s'} C(s,s')U_{\mathrm{I}}(s,s')\vec{{\mu}}(s')U^\dagger_{\mathrm{I}}(s,s')\cdot \vec{\sigma} .
\eeq


\section{Necessary convergence criterion}
\label{app:F}

\subsection{General criterion}

Assuming $[\rho_B,H_B]=0$ the correlation function becomes homogeneous in time, so we use the shorthand notation $C\pqty{x-y} = C\pqty{x-y,0} = C\pqty{x,y}$. We define the following quantities to bound the error of the expansion:
\begin{equation}
\label{eq:correlation_integral}
	\tau^{\pqty{n}}_B = t_f^n\int_0^\infty \dd{s} s^{n-1}\abs{C(s)} ,
\end{equation}
and denote $\tau_B^{(1)} \equiv \tau_B$, which has a natural interpretation as the bath correlation time~\cite{ABLZ:12-SI}.

Note that
\begin{equation}
	\norm{\mathcal{K}_2(s) [\tilde{\rho}_S\otimes\rho_B]} \leq c_2 \kappa^2 \int_0^s \dd{s'}\abs{C\pqty{s'}} \leq c_2\kappa^2\tau_B /t_f ,
\end{equation}
where $c_2=O(1)$ is a constant arising from the number of terms in the TCL2 double commutator expression~\eqref{eq:tcl2_generator}.
We can similarly estimate the magnitude of the TCL4 terms:
\begin{align}
	&\norm{\mathcal{K}_4(s) [\tilde{\rho}_S\otimes\rho_B]} \notag \\
	&\leq c_4 \kappa^4\int_0^s\int_0^{s_1}\int_0^{s_2}\abs{C\pqty{s-s_2}}\abs{C\pqty{s_1-s_3}}\dd{s_1}\dd{s_2}\dd{s_3} \notag \\
	&\quad +c_4'\kappa^4 \int_0^s\int_0^{s_1}\int_0^{s_2}\abs{C\pqty{s-s_3}}\abs{C\pqty{s_1-s_2}}\dd{s_1}\dd{s_2}\dd{s_3} \label{eq:tcl4}
\end{align}
where $c_4, c_4'=O(1)$ are constants arising from the number of terms in the TCL4 sum over multiple commutators and triple integral. We can bound the two integrals in Eq.~\eqref{eq:tcl4} as  follows. Considering the first expression, we first make a change of variables as
\begin{equation}
	x_1 = s-s_3 \quad x_2 = s_1 - s_3 \quad x_3 = s_2 - s_3 .
\end{equation}
Because $1 \ge s \ge s_1 \ge s_2 \ge s_3$, the new integration limits can be obtained as
\bes\begin{align}
	&s_3 \ge 0 \implies s \ge x_1 &&s \ge s_1 \implies x_1 \ge x_2 \\
	&s_1 \ge s_2 \implies x_2 \ge x_3  &&s_2\ge s_3 \implies x_3\ge 0
\end{align}\ees
which is $s\ge x_1 \ge x_2 \ge x_3\ge 0$. The Jacobian $\abs{\det J}=1$ in this case and the new integral becomes
\bes
\begin{align}
	&\int_0^s\dd{x_1}\int_0^{x_1}\dd{x_2}\int_0^{x_2}\dd{x_3} \abs{C\pqty{x_1-x_3}}\abs{C\pqty{x_2}} \\
	&\quad \leq \int_0^s\dd{x_1}\int_0^{x_1}\dd{x_2}\int_0^{x_1}\dd{x_3} \abs{C\pqty{x_1-x_3}}\abs{C\pqty{x_2}} \\
	&\quad \leq \int_0^s\dd{x_1}\int_0^{x_1}\dd{x_3} \abs{C\pqty{x_1-x_3}}\int_0^{x_1}  \dd{x_2}\abs{C\pqty{x_2}}  \\
	&\quad \leq \int_0^s\dd{x_1}\int_0^{x_1}\dd{x_3}\abs{C\pqty{x_1-x_3}} \frac{\tau_B}{t_f} .
\end{align}
\ees
Now we make another change of variables, with
\begin{equation}
	v = x_1 - x_3 \quad u = x_1 + x_3 .
\end{equation}
The new integration limits can be obtained by
\bes
\begin{align}
	x_3 &\ge 0 \implies u  \ge v \quad s \ge x_1 \implies 2s-v \ge u \\
	x_1 &\ge x_3 \ge 0 \implies 0 \le v \le x_1 .
\end{align}
\ees
The first line means that $2s-v \ge u  \ge v$, while the second line gives $0\le v \le s$ since $0\le x_1 \le s$. Thus:
\bes
\begin{align}
	&\int_0^s \dd{x_1}\int_0^{x_1}\dd{x_3}\abs{C\pqty{x_1-x_3}}  \\
	&\quad = \int_0^s \dd{v} \abs{C\pqty{v}} \int_{v}^{2s-v} \dd{u}  \abs{\det J} \\
	&\quad = \int_0^s\frac{1}{2}\pqty{2s-2v}\abs{C\pqty{v}}\dd{v} \\
	&\quad \le \int_0^\infty s \abs{C\pqty{v}}\dd{v} \leq \frac{\tau_B}{t_f} ,
\end{align}
\ees
where in the last inequality we used $s\le 1$.

The same can be done for the second integral in Eq.~\eqref{eq:tcl4}:
\bes
\begin{align}
	&\int_0^s \dd{s_1} \int_0^{s_1} \dd{s_2} \int_0^{s_2} \dd{s_3} \abs{C\pqty{s-s_3}}\abs{C\pqty{s_1-s_2}} \\
	&\quad =\int_0^s \dd{x_1}\int_0^{x_1}\dd{x_2}\int_0^{x_2}\dd{x_3}\abs{C\pqty{x_1}}\abs{C\pqty{x_2-x_3}} \\
	&\quad \leq \int_0^s \dd{x_2}\int_0^{x_2} \dd{x_3}\abs{C\pqty{x_2-x_3}} \int_0^{s} \dd{x_1} \abs{C\pqty{x_1}} \\
	&\quad \leq 
	\left({{\tau_B}}/{t_f}\right)^2.
\end{align}
\ees
Combining these two bounds thus finally yields:
\begin{align}
	\norm{\mathcal{K}_4(s) [\tilde{\rho}_S\otimes\rho_B]} 
	\leq \pqty{c_4+c_4'}\kappa^4 \left({{\tau_B}}/{t_f}\right)^2
\end{align}
In particular, to ensure the validity of the TCL2 approximation it should be the case that the TCL4 term is much smaller than TCL2, i.e.:
\beq
\label{eq:tcl2_valid}
g^2 t_f \tau_B < \frac{c_2}{c_4+c'_4} \quad \textrm{or} \quad g^2 t_f \tau_B \ll 1.
\eeq
%


\subsection{Ohmic bath case}

Let us assume a spin-boson noise model, for which 
\bes
\begin{align}
	H_{SB} &= 
	g Y\otimes \sum_k\pqty{\xi_k^* b_k^\dagger + \xi_k b_k} \\
	H_B &= \sum_k \omega_k b_k^\dagger b_k ,
\end{align}
\ees
where $b_k$ is a bosonic annihilation operator for mode $k$ with frequency $\omega_k$, and $g_k = g\xi_k$ is the associated system-bath coupling strength, where $\xi_k$ is dimensionless and $g$ has units of energy.
A standard approach is to introduce a spectral density such that $\abs{g_k}^2 \mapsto J\pqty{\omega}d\omega$. For an Ohmic bath we have 
\begin{equation}
	J\pqty{\omega} = \eta \omega e^{-\omega/\omega_c} ,
\end{equation}
where $\eta$ is a parameter with dimensions of time squared.
After transforming to the bath interaction picture and replacing $t$ by $s=t/t_f$ to arrive at $\tilde{H}_{SB}(s)$, the bath correlation function for the Ohmic spectral density is
\begin{align}
	&C(s)= \int_0^\infty\dd{\omega} \eta\omega e^{-\omega/\omega_c}\times  \\
	&\qquad \qquad\qquad \pqty{\coth(\frac{\beta\omega}{2})\cos(\omega s t_f )-i\sin(\omega s t_f)}, \notag
\end{align}
an integral which may be evaluated explicitly in terms of the Polygamma function~\cite{ABLZ:12-SI}. 
In particular, for large $\beta \omega_c$ and $t_f/\beta$, the correlation function can be expanded as
\begin{align}
	C(s) &= \frac{\eta}{\beta^2}\Bigg(-4\pi^2 e^{-s t_f/\tau_B}+\frac{1}{\pqty{st_f/\tau_M}}\notag \\
	&\quad +O\pqty{e^{-2st_f/\tau_B},\pqty{st_f}^{-3}} \Bigg) .
	\label{eq:C-Ohmic}
\end{align}
This form indicates a transition from a Markovian regime of purely exponential decay with a timescale of $\tau_B \overset{\omega_c\to\infty}{\to} {\beta}/({2\pi})$, followed by a non-Markovian regime of power-law decay with a timescale of $\tau_M = \sqrt{2\beta/\omega_c}$. The transition occurs at a time $\tau_{\mathrm{tr}} \approx \beta\ln(\beta\omega_c)$~\cite{ABLZ:12-SI}.
In the Markovian limit $\omega_c\to \infty$ we may thus replace Eq.~\eqref{eq:C-Ohmic} by
\beq
|C(s)| = \eta \left(\frac{2\pi}{\beta}\right)^2 e^{-2\pi s t_f/\beta} ,
\eeq
and hence the correlation function integral of Eq.~\eqref{eq:correlation_integral} becomes
\beq
	\int_0^\infty \dd{s} \abs{C(s)} =     \frac{\eta}{2\pi t_f \beta}  ,
\eeq
which replaces every factor of $\tau_B/t_f$ arising from the same integral in the bounds in the previous subsection. In particular, we now have the necessary condition $\norm{\mathcal{K}_4(s) [\tilde{\rho}_S\otimes\rho_B]} \leq \pqty{c_4+c_4'}\alpha^4 \left( \eta /({2\pi t_f \beta})\right)^2 < \norm{\mathcal{K}_2(s) [\tilde{\rho}_S\otimes\rho_B]} \leq c_2\alpha^2 \eta   /({2\pi t_f \beta})$.
Eq.~\eqref{eq:tcl2_valid} can thus be rewritten in the Markovian Ohmic case as
\begin{equation}
	\frac{g^2 \eta t_f}{\beta } < \frac{2\pi c_2}{\pqty{c_4+c_4'}}    \quad \textrm{or} \quad \frac{g^2 \eta t_f}{\beta } \ll 1 .
	\label{eq:bound-Ohmic}
\end{equation}
For finite $\omega_c$, one can refine this bound by replacing Eq.~\eqref{eq:correlation_integral} with
\begin{equation}
	\tau_B = t_f\left(\int_0^{s_{tr}}\dd{s}\abs{C(s)}+\int_{s_{tr}}^{\infty}\dd{s}\abs{C(s)}\right) ,
\end{equation}
where $s_{\mathrm{tr}} = \tau_{\mathrm{tr}}/t_f$. For our purposes the bound~\eqref{eq:bound-Ohmic} suffices and is satisfied in all the numerical results presented in the main text. Namely, we have $\frac{g^2 \eta t_f}{\beta } \leq 0.16$.


\section{Rotating wave approximation}
\label{app:G}

Let
\begin{equation}
	\Gamma(\omega) = \int_0^\infty \dd{t} e^{i\omega t} C(t) = t_f \Gamma_s(\omega t_f)
\end{equation}
be the one-sided Fourier transform of the bath correlation function,
where
\begin{equation}
	\Gamma_s\pqty{\omega}\equiv\int_0^\infty\dd{s}e^{i\omega s}C(s) =\frac{1}{2}\gamma_s\pqty{\omega} + iS_s\pqty{\omega} ,
\end{equation}
and where $\gamma_s\pqty{\omega}/2$ and $S_s\pqty{\omega}$ are the real and imaginary parts of $\Gamma_s\pqty{\omega}$. Explicitly~\cite{Breuer:book}:
\begin{subequations}
\begin{align}
\gamma_s(\omega) &= \int_{-\infty}^\infty e^{i\omega s} C(s) ds \\
S_s(\omega) &= \frac{1}{2\pi} \int_{-\infty}^\infty \gamma(\omega') \mathcal{P}\left(\frac{1}{\omega-\omega'}\right)d \omega' \ .
\end{align}
\end{subequations}
Here $\mathcal{P}$ denotes the Cauchy principal value, and the $s$ subscript is a reminder that $t_f$ has been factored out.

To perform the rotating wave approximation, let us first define the eigenspace projection operator of $H_{\mathrm{I}}(s)$ as
\begin{equation}
    \Pi\pqty{\epsilon(s)} = \dyad{\epsilon(s)} ,
\end{equation}
where $\ket{\epsilon(s)}$ is an eigenstate of $H_{\mathrm{I}}(s)$ with instantaneous energy $\epsilon(s)$.
We can then define the operator
\begin{equation}
\label{eq:a-op}
	A\pqty{\omega(s)} \equiv \sum_{\epsilon'(s)-\epsilon(s)=\omega(s)}\Pi\pqty{\epsilon(s)}\big[\vec{\mu}(s)\cdot\vec{\sigma}\big]\Pi\pqty{\epsilon'(s)} ,
\end{equation}
where 
\beq
\omega(s) \in \Bqty{0,\pm\dot{\theta}(s)}
\label{eq:Bohr}
\eeq 
is the dimensionless Bohr frequency, and the sum is over all pairs $\epsilon(s),\epsilon'(s)$ subject to the constraint $\epsilon'(s)-\epsilon(s)=\omega(s)$. The interaction picture master equation~\eqref{eq:IPME} can then be written to second order, with the TCL2 generator~\eqref{eq:tcl2_generator} as
\begin{align*}
	\dot{\tilde{\rho}}_S &=\int_0^s\dd{s'}\Tr_B\comm{\tilde{H}_{\mathrm{tot}}(s)}{\comm{\tilde{H}_{\mathrm{tot}}\pqty{s'}}{\tilde{\rho}_S(s)\otimes\rho_B}} \\
	&=\kappa^2\sum_{\omega,\omega'}e^{i\pqty{\omega'-\omega}s}\Gamma\pqty{\omega}\bigg(A\pqty{\omega}\tilde{\rho}_S A^\dagger\pqty{\omega'}\\
	&\qquad\qquad\qquad-A\pqty{\omega'}A\pqty{\omega}\tilde{\rho}_S\bigg) +\textrm{h.c.} \numberthis 
	\label{eq:freq_TCL2}
\end{align*}
To obtain this master equation, we apply the standard Markovian approximation: change the integration variable $s' \mapsto s-s'$ and replace the upper limit with $\infty$. The RWA consists of neglecting terms in Eq.~\eqref{eq:freq_TCL2} for which $\omega' \neq \omega$. A necessary condition for the validity of the RWA is~\cite{Lidar:2019aa}:
\begin{equation}
	1/\tau_B < \min_{\omega \neq \omega'}\abs{\omega - \omega'}\ ,
\end{equation}
which, unfortunately, is not always satisfied for the two-step Gaussian schedule~\eqref{eq:sumofG} because [recall Eq.~\eqref{eq:Bohr}]
\begin{equation}
	\min_{\omega \neq \omega'}\abs{\omega - \omega'} = \dot{\theta}(s) \approx 0
\end{equation}
for $s$ outside the Gaussian pulse region.

Nevertheless, the RWA results in the interaction picture adiabatic Markovian master equation in Lindblad form~\cite{ABLZ:12-SI}:
\begin{equation}
\label{eq:rwa_interaction}
	\dot{\tilde{\rho}}_S = -i\comm{H_{\mathrm{LS}}}{\tilde{\rho}_S} + \mathcal{D}\pqty{\tilde{\rho}_S} ,
\end{equation}
where 
\begin{equation}
	H_{\mathrm{LS}} = \kappa^2 \sum_\omega S_s\pqty{\omega}A^\dagger\pqty{\omega}A\pqty{\omega}
\end{equation}
is the Lamb shift, 
and 
\begin{align}
	&\mathcal{D}\pqty{\tilde{\rho}_S}=\kappa^2 \sum_{\omega}\gamma_s\pqty{\omega}\bigg((A\pqty{\omega}\tilde{\rho}_SA^\dagger\pqty{\omega} \notag\\
	&\qquad\qquad\qquad  -\frac{1}{2}\acomm{A^\dagger\pqty{\omega}A\pqty{\omega}}{\tilde{\rho}_S}\bigg)
\end{align}
is the dissipator. 

We can explicitly calculate $A\pqty{\omega(s)}$. First, recalling that
$H_{\mathrm{I}}(\tau) =  \frac{1}{2}\dv{\theta}{\tau} U_0^\dagger(\tau)X U_0(\tau)$ 
[Eq.~\eqref{eq:intp}], 
we realize that the eigenvalues and eigenvectors of $H_{\mathrm{I}}(s)$ can be written as
\begin{align}
	&\epsilon_\pm(s) = \pm \frac{1}{2}\dot{\theta}(s) \quad &&\ket{\epsilon_\pm(s)}=U_0^\dagger(s)\ket{\pm}	 \ .
\end{align}
Also, from the sequence of transformations leading to Eq.~\eqref{eq:interaction-h}, the interaction terms have the form
\begin{equation}
	\vec{\mu}(s)\cdot\vec{\sigma} = U_0^\dagger(s)\Big(\cos\theta(s) Y+\sin\theta(s) Z\Big)U_0(s)\ .
\end{equation}
Substituting these expressions back into Eq.~\eqref{eq:a-op}, we obtain
\bes
\begin{align}
	A(0) &= 0 \\
	A\pqty{\dot{\theta}(s)}&= -ie^{i\theta}\dyad{\epsilon_-(s)}{\epsilon_+(s)} \\
	A\pqty{-\dot{\theta}(s)}&= ie^{-i\theta}\dyad{\epsilon_+(s)}{\epsilon_-(s)}\ .
\end{align}
\ees
After undoing the interaction picture transformation with respect to $H_{\mathrm{I}}(s)$ 
and ignoring the phase factors in the $A\pqty{\omega}$ operators, we obtain the  Schr\"odinger picture master equation, namely Eqs.~\eqref{eq:TCL2_RWA}-\eqref{eq:H_LS} given in the main text.
In deriving this result we made use of the Kubo-Martin-Schwinger (KMS) condition~\cite{Breuer:book}
\beq
\gamma(-\Delta) = e^{-\beta\Delta}\gamma(\Delta)\ ,
\eeq
where $\Delta$ is the dimensionless Bohr frequency in units of $1/t_f$:
\begin{equation}
	\Delta(s) = \omega(s) / t_f \ .
\end{equation}
%

\section{Derivation of the semi-empirical Eq.~\eqref{eq:empirical_1}}
\label{app:H}
The semi-empirical formula~\eqref{eq:empirical_1} can be derived directly from Eq.~\eqref{eq:rwa_interaction}. Let us first write Eq.~\eqref{eq:rwa_interaction} in terms of the quantities defined in Eq.~\eqref{eq:22b}:
\begin{align}
\label{eq:h1}
    \dot{\tilde{\rho}}_S  = &-i[H_{LS}, \tilde{\rho}_S] \notag \\
    &-t_f \gamma_d \big(\tilde{\rho}_{+-}\dyad{\epsilon_+}{\epsilon_-}+\tilde{\rho}_{-+}\dyad{\epsilon_-}{\epsilon_+} \big) \\
    &+t_f \gamma_t (\tilde{\rho}_{++}-e^{-\beta \Delta}\tilde{\rho}_{--})(\dyad{\epsilon_-}{\epsilon_-}-\dyad{\epsilon_+}{\epsilon_+}) \notag \ .
\end{align}
We now follow the steps in Ref.~\cite{Albash:2015nx} to obtain the solution in this interaction picture. Eq.~\eqref{eq:h1} can be split into two decoupled ordinary differential equations:
\bes
\label{eq:split}
\begin{align}
    \dv{\tilde{\rho}_{--}}{s} &= -\dv{\tilde{\rho}_{++}}{s} \notag \\
    &=\bqty{\mathcal{F}_+\pqty{s}\tilde{\rho}_{++}-\mathcal{F}_-\pqty{s}\tilde{\rho}_{--}} \\
    \dv{\tilde{\rho}_{+-}}{s} &= \dv{\tilde{\rho}_{-+}^*}{s} \notag \\
    &= -\bqty{i\Omega(s) + \Sigma(s)}\tilde{\rho}_{+-}\ ,
\end{align}
\ees
where
\bes
\begin{align}
    \mathcal{F}_{+}(s) &= g^2 t_f\gamma_t(s) \\
    \mathcal{F}_{-}(s) &= g^2 t_f\gamma_t(s)e^{-\beta \Delta(s)}\ ,
\end{align}
\ees
and
\bes
\begin{align}
    \Omega(s) &= g^2 t_f(S(\Delta(s))-S(-\Delta(s))) \\
    \Sigma(s) &= g^2 t_f\gamma_d(s) \ .
\end{align}
\ees
Additionally, the KMS condition allows us to write $\gamma_d(s)$ in terms of $\mathcal{F}_+(s)$
\begin{equation}
    \mathcal{F}_+(s)(1+e^{-\beta \Delta(s)}) = 2 g^2 t_f \gamma_d(s)\ .
\end{equation}
The solution of Eqs.~\eqref{eq:split} is given by:
\bes
\label{eq:solution_pp}
\begin{align}
    \rho_{--}(s) &= \exp[-2t_fg^2\int_0^s \dd{s'}\gamma_d(s')] \Bigg\{\rho_{--}(0) \label{eq:solution_mm} \\
    &\quad+\int_0^s\dd{s'}\mathcal{F}_+(s') \exp[2t_fg^2\int_0^{s'}\dd{s''}\gamma_d(s'')] \Bigg\} \notag \\
    \rho_{+-}(s) &= \exp{-\int_0^s\dd{s'}[i\Omega(s') +t_fg^2 \gamma_d(s')]}\rho_{+-}(0)  \\
    \rho_{++}(s) &= 1-\rho_{--}(s) \\
    \rho_{-+}(s) &= \rho^*_{+-}(s)\ ,
\end{align}
\ees
where the initial conditions are:
\begin{align}
    \rho_{ij}(0) = \frac{1}{2}\ , \quad i,j \in \{+,-\} .
\end{align}

The next step is to move back to Schr\"odinger picture
\begin{equation}
    \rho_S(t) = U_{\mathrm{I}}(t) \tilde{\rho}_S(t) U_{\mathrm{I}}^\dagger(t)\ ,
\end{equation}
and write the open system ground state probability in terms of  $\tilde{\rho}_S$:
\bes
\label{eq:PG'}
\begin{align}
	P_G'(t_f) &=\ev{\rho(t_f)}{0}  
	\\&= \ev{U_{\mathrm{I}}(t_f) \tilde{\rho}(t_f) U_{\mathrm{I}}^\dagger(t_f)}{0} \\
    &=\sum_{i,j \in \{+,-\}}\rho_{ij}\langle 0 \dyad{\chi_i}{\chi_j} 0 \rangle\ ,
\end{align}
\ees
where
\begin{equation}
    \ket{\chi_i(t_f)} = U_{\mathrm{I}}(t_f)\ket{\epsilon_i(t_f)} =U_{\mathrm{I}}(t_f)U_0^\dagger(t_f)\ket{i} \ .
\end{equation}
For simplicity, we further denote $U^a(t) = U_{\mathrm{I}}(t)U_0^\dagger(t)$, whose elements can be related to those of $U_{\mathrm{I}}(t)$ in the $\{\ket{0},\ket{1}\}$ basis:
\begin{equation}
	U_{kl}^a(t) = \mel{k}{U_{\mathrm{I}}(t)U_0^\dagger(t)}{l} = e^{(-1)^{l}i\phi(t)}\mel{k}{U_{\mathrm{I}}(t)}{l} \ ,
\end{equation}
where $k,l \in \{0,1\}$ and $\phi(t) = -E_0 t / 2$. Then:
\bes
\begin{align}
    \langle 0 \dyad{\chi_+}0 \rangle &= \frac{1}{2}(\abs{U_{00}^a}^2+U_{00}^aU_{01}^{a*}+U_{01}^aU_{00}^{a*}+\abs{U_{01}^a}^2) \\
    \langle 0 \dyad{\chi_-}0 \rangle &=\frac{1}{2}(\abs{U_{00}^a}^2-U_{00}^aU_{01}^{a*}-U_{01}^aU_{00}^{a*}+\abs{U_{01}^a}^2) \\
    \langle 0 \dyad{\chi_+}{\chi_-}0 \rangle &=\frac{1}{2}(\abs{U_{00}^a}^2-U_{00}^aU_{01}^{a*}+U_{01}^aU_{00}^{a*}-\abs{U_{01}^a}^2) \\
    \langle 0 \dyad{\chi_-}{\chi_+}0 \rangle &=\frac{1}{2}(\abs{U_{00}^a}^2+U_{00}^aU_{01}^{a*}-U_{01}^aU_{00}^{a*}-\abs{U_{01}^a}^2) \ .
\end{align}
\ees
Because $U_{\mathrm{I}}(t)$ is the closed system unitary, we have 
\begin{equation}
    \abs{U_{00}^a(t_f)}^2 = \abs{\ev{U_{\mathrm{I}}(t_f)}{0}}^2 = P_G(t_f)\ ,
\end{equation}
and
\begin{equation}
	\abs{U^a_{00}}^2 + \abs{U^a_{01}}^2 = 1 \ .
\end{equation}
Eq.~\eqref{eq:PG'} becomes:
\bes
\label{eq:PG'2}
\begin{align}
    P'_G(t_f) &= \frac{1}{2} + (\rho_{+-}(t_f)+\rho_{-+}(t_f))(P_G(t_f)-\frac{1}{2}) \label{eq:rwa_2_empirical}\\
    &\quad+(\rho_{++}(t_f)-\rho_{--}(t_f))\Re(U^a_{00}U_{01}^{a*}) \label{eq:empirical-simplify}\\
    &\quad+i(\rho_{-+}(t_f)-\rho_{+-}(t_f))\Im (U^a_{00}U_{01}^{a*})\ .
\end{align}
\ees
This result is exact and corresponds to the numerical solution in the TCL2 case shown in Fig.~\ref{fig:redfield_rwa}.

We now make two additional approximations in order to arrive at a simpler expression. First, we ignore the Lamb shift term $\Omega(s)$ in Eqs.~\eqref{eq:solution_pp}, which leads to:
\bes
\begin{align}
    \rho_{+-}(t_f) + \rho_{-+}(t_f) &\approx \exp{-g^2 t_f \int_0^1 \dd{s} \gamma_d(s)} \\
     \rho_{+-}(t_f) - \rho_{-+}(t_f) &\approx 0\ .
\end{align}
\ees
Second, we substitute the solution given in Eqs.~\eqref{eq:solution_pp} into line~\eqref{eq:empirical-simplify}:
\bes
\begin{align}
    &(\rho_{++}(t_f)-\rho_{--}(t_f))\Re(U^a_{00}U_{01}^{a*}) = \notag \\
	&\qquad \Re(U^a_{00}U_{01}^{a*})\Bigg\{1-2e^{-2t_fg^2\int_0^1 \dd{s'}\gamma_d(s')}\bigg[ \frac{1}{2}\notag \\
	&\qquad+t_fg^2\int_0^1\dd{s'}\gamma_t(s') e^{2 g^2 t_f\int_0^{s'}\dd{s''} \gamma_d(s'')} \bigg] \Bigg\} \\
	&\qquad\approx (1-2\frac{1}{2})\Re(U^a_{00}U_{01}^{a*}) = 0 \ ,
\end{align}
\ees
where in the last line we used the weak coupling assumption, $g^2 t_f \ll 1$.

With these two approximations, Eq.~\eqref{eq:PG'2} becomes the semi-empirical formula~\eqref{eq:empirical_1} with $P_E(0)=1/2$. We note that it is well known that for time-independent Lindbladians the RWA master equation has the Gibbs state as its steady state~\cite{Breuer:book}. We do not recover this result for the time-dependent case. Rather, we find that the time-dependent Redfield master equation (TCL2) converges to the Gibbs state $P_E(\beta) = \frac{e^{\beta E_0/2}}{Z}$, but with a temperature that differs from that of the bath state, as illustrated in Fig.~\ref{fig:redfield_rwa}.



%

\end{document}